\documentclass[aps,prd,showpacs,nofootinbib]{revtex4-1}
\usepackage[latin1]{inputenc}
\usepackage[T1]{fontenc}
\usepackage{ifpdf}
\ifpdf
\usepackage[pdftex]{graphicx}
\else
\usepackage[dvips]{graphicx}
\fi

\usepackage{amssymb}
\usepackage{amsmath}
\usepackage{verbatim}

\usepackage{color}
\usepackage{revsymb4-1}
\usepackage[usenames,dvipsnames,svgnames,table]{xcolor}
\usepackage{braket}
\ifpdf
\usepackage[pdftex]{hyperref}
\else
\usepackage[dvips]{hyperref}
\fi
\def \Kla#1{\left( #1 \right)}
\def \Klb#1{\left[ #1 \right]}
\def \KlB#1{\left| #1 \right|}

\def \Unit#1{\,\hbox{\text{#1}}}

\def \UnitMu{\Unit{$\mu$m}}
\def \etal {et al$.$}
\def \AiryAi{\mathop{\rm Ai}\nolimits}
\def \AiryAP{\mathop{\rm Ai}^\prime}
\def \mn{m_{\rm n}}
\def \mun{{\mu_{\rm n}}}
\def \eH#1{{\rm e}^{#1}}
\def \Spm{{s_\pm}}
\def \Dmag{{D_\pm}}
\def \zstep{{\Delta z_{\rm step}}}
\def \AfterStep{\hbox{after step}}
\def \QE{\widetilde{E}}

\begin{document}


\title{Frequency shifts in gravitational resonance spectroscopy}

\date{Nov 15, 2014}

\author{S. Bae\ss{}ler}
\affiliation{Physics Department, University of Virginia\\
382 McCormick Road, Charlottesville, VA 22904, U.S.A.
}%
\affiliation{Oak Ridge National Laboratory\\
1 Bethel Valley Road, Oak Ridge, TN 37831, U.S.A
}%
\email[Corresponding author. ]{E-mail: baessler@virginia.edu}

\author{V.V. Nesvizhevsky}
\affiliation{%
Institut Laue-Langevin\\
71 avenue des Martyrs, 38042 Grenoble, France
}%

\author{G. Pignol}%
\affiliation{%
LPSC, Université Grenoble-Alpes, CNRS/IN2P3\\
53 rue des Martyrs, 38026 Grenoble, France
}%

\author{K.V. Protasov}%
\affiliation{%
LPSC, Université Grenoble-Alpes, CNRS/IN2P3\\
53 rue des Martyrs, 38026 Grenoble, France
}%

\author{D. Rebreyend}%
\affiliation{%
LPSC, Université Grenoble-Alpes, CNRS/IN2P3\\
53 rue des Martyrs, 38026 Grenoble, France
}%

\author{E.A. Kupriyanova}%
\affiliation{%
P.N. Lebedev Physical Institute\\
53 Leninskii pr., 119991 Moscow, Russia 
}%

\author{A.Yu. Voronin}%
\affiliation{%
P.N. Lebedev Physical Institute\\
53 Leninskii pr., 119991 Moscow, Russia 
}%

\begin{abstract}
Quantum states of ultracold neutrons in the gravitational field are to be characterized through gravitational resonance spectroscopy. This paper discusses systematic effects that appear in the spectroscopic measurements. The discussed frequency shifts, which we call Stern-Gerlach shift, interference shift, and spectator state shift, appear in conceivable measurement schemes and have general importance. These shifts have to be taken into account in precision experiments.
\end{abstract}

\pacs{04.80.Cc,04.90.+e,07.05.Fb,29.90.+r}

\maketitle

\section{Introduction}
Gravitationally bound quantum states of ultracold neutrons have been discovered in the last decade \cite{NesNat02,NesPRD03,NesEPJC05,BaeJPG09,NesUsp10}. These quantum states are formed if ultracold neutrons with very little energy are brought on top of a horizontal mirror that has essentially infinitely high potential for the neutrons. The early experiments were sensitive to the shape of the wave function of these quantum states. Higher precision is achievable in gravitational resonance spectroscopy, that is, if the differences in energy of the lowest gravitational quantum states are detected \cite{NesBook05}. The interaction that couples different quantum states can be a vibration of the bottom mirror, or an oscillating magnetic field. The former is planned in the QuBounce project, and has shown first results \cite{JenNat11, JenPRL14}. The latter is planned for the GRANIT project \cite{Kre09,Pig09,Bae11a}. The ultimate goal of the GRANIT collaboration is to perform gravitational resonance spectroscopy on ultra-cold neutrons trapped in quantum states \cite{Nes07}. The length of the observation time would allow a very high precision in the measurement of energy differences between those quantum states.

The motivation for these measurements comes from the sensitivity of the gravitationally bound quantum states to extra short-range interactions (See \cite{Ant11} and references therein) that might be spin-dependent, or from the possibility to detect Chameleons \cite{Bra11}. Furthermore, the study of gravitationally bound quantum states allows for a test of the weak equivalence principle that is probably the only one that uses bound quantum states, albeit its sensitivity cannot compete with tests using free-falling cold atoms \cite{Pet99} or macroscopic bodies \cite{Adel09}.


The paper is organized as follows: In section 2, we introduce spectroscopy of quantum mechanical bound states, and describe the setups that have been proposed to detect them. In section 3, we introduce the Stern-Gerlach shift that gives the largest shift in the setups that use a spatially varying magnetic field. In section 4, we introduce the spectator state shift, which is present in all measurement schemes. In section 5, we discuss the interference shift that is washed out for all measurement schemes that use a time-dependent perturbation. 

\section{Spectroscopy of quantum-mechanical bound states}

The one-dimensional quantum states of ultracold neutrons in the gravitational field and above a perfect reflecting horizontal mirror at $z=0$ are given by $\psi_m(z) = A_m\AiryAi((z-z_m)/z_0)$ (for $z>0$), where $\AiryAi(z)$ is the Airy function, $z_0=\Kla{\hbar^2/2\mn^2 g}^{1/3}\sim 5.87$\UnitMu{} is the characteristic length scale of the problem,
 and $z_m$ is the $m^{\text{th}}$ solution of $\AiryAi(-z_m/z_0)=0$. The normalization factor can be written as
$A_m=1/\KlB{\AiryAP(-z_m/z_0)}$. 
Energy eigenvalues of the lowest quantum states are very small, $E_m=\mn gz_m$, and are on the order of peV. The height of the lowest quantum states is of the order of tens of micrometers.



A generic setup to perform spectroscopy consists of the following steps:
\begin{itemize}
\item[(1.)] Preparation of an initial quantum state.
\item[(2.)] Transition to the final quantum state using a periodic potential. The transition is only efficient if the frequency of the periodic potential matches the energy difference between initial and final state. 
\item[(3.)] State analysis and detection.
\end{itemize}

There have been several proposals to achieve gravitational resonance spectroscopy in flow-through mode, that is, using ultracold neutrons that traverse a setup:
\begin{itemize}
\item[(A)] {\bf Magnetic transitions in DC mode:} A generic setup is shown in Fig. \ref{fig:GRANITSetup}.
The downward-going step in the bottom mirror was proposed first in ref. \cite{Nes00} as a tool to depopulate the ground state. A set of parallel wires with an oscillating current pattern generates a magnetic field that rotates multiple times; a possible choice is shown in the picture. In addition, there is a vertical holding field. This resulting rotating magnetic field gradient induces transitions between quantum states if the rotation frequency (as seen by the neutron) matches the energy difference of the quantum states. A ground state filter made from the bottom mirror and a scatterer with a rough bottom surface that leaves an open slit height of about $z_{\rm slit} \sim 25$\UnitMu{} accepts ground state neutrons and rejects neutrons  in higher quantum states. Earlier experiments \cite{NesNat02,NesPRD03,NesEPJC05} demonstrate the operation of this filter. The resonance condition in the transition region can only be fulfilled for neutrons with a certain velocity in forward direction. After a short free-fall region, neutrons fall a height which is given by their velocity component in forward direction. The neutron detector at the end of the free-fall region is position sensitive in the vertical direction, and a given vertical coordinate corresponds to a given forward velocity. This setup has been proposed earlier in \cite{Kre09,Bae11a}. 
\begin{figure}[htb]
\includegraphics{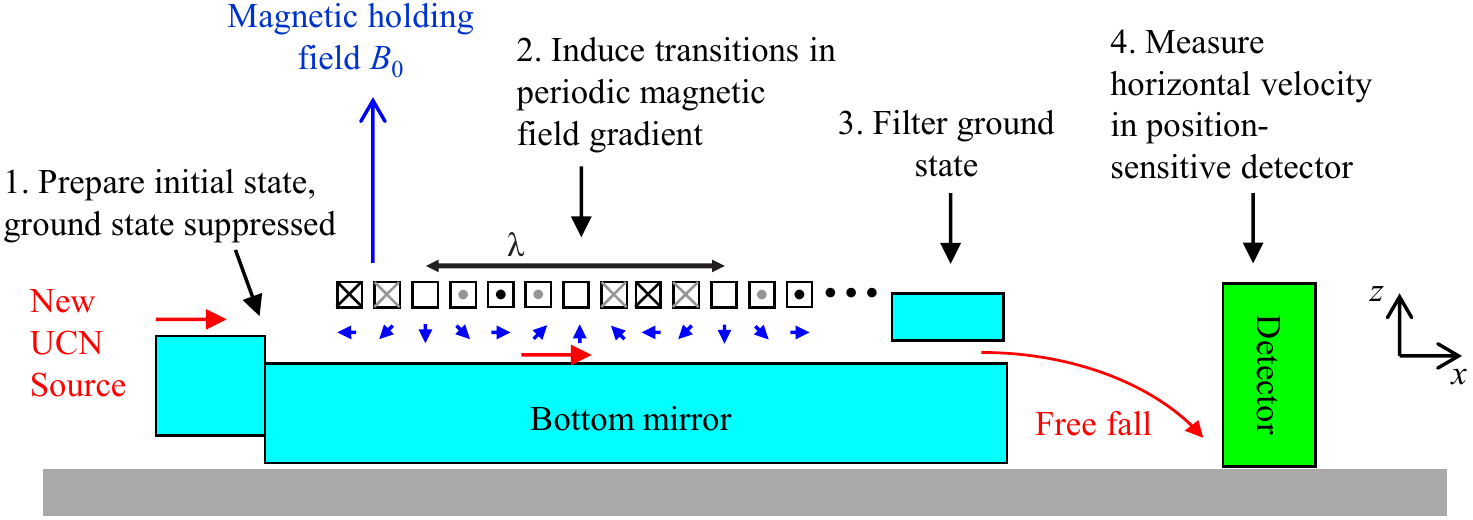}
\caption{Sketch of the flow-through setup. Neutrons enter from the left. They go through the state preparation region (1.), transition region (2.), state analysis region (3.) and detection (4.). The choice of parameters used in the simulations in this paper is a current pattern $I_0, I_0/\sqrt{2}, 0, -I_0/\sqrt{2}, -I_0, -I_0/\sqrt{2}, 0, \dots$ that repeats 24 times, and a periodicity of $\lambda = 1$\Unit{cm}.}
\label{fig:GRANITSetup} 
\end{figure}

\item[(B)] {\bf Magnetic transitions in AC mode:} The setup is similar to the one in DC-mode, but here, the current in the wires is oscillating in time. The vertical holding field is replaced by a very small holding field along the wire direction, which serves only to avoid spin flip transitions at the zero crossings of the current. The neutron spin is rotating with the magnetic field, and the temporal oscillation of the field magnitude leads to transition if the frequency matches an energy difference between quantum states. The neutron velocity matters only as it influences the time the neutron spends in the transition region, and therefore the neutron detector does not need to be position-sensitive. This setup has been described in more detail by Pignol \etal{} \cite{Pign14}.

\item[(C)] {\bf Mechanically induced transitions:} An alternate way to produce an oscillating potential that induces quantum state transitions is to vibrate the bottom mirror. A proposal has been made in \cite{Abe10} for such an experiment, with the additional feature that the long transition region is split into two short ones with a long bottom mirror in between, similar to Ramsey spectroscopy in atomic physics. Before the transition region, a state filter selects only the ground state. Behind the transition region, another state filter selects the ground state again, and a detector behind the setup would count only the neutrons for which no transition has occurred. Results have been reported from a more condensed version of this setup by Jenke \etal{} \cite{JenNat11, JenPRL14}, where the three regions have not been separated; we are not discussing this complication here.

\end{itemize}

The Hamiltonian for an UCN above a horizontal mirror at $z=0$ in the transition region equations is
\begin{equation}
H\Ket{\psi(t)}=H_0\Ket{\psi(t)}+V(t)\Ket{\psi(t)}=i\hbar\frac{\partial}{\partial t}\Ket{\psi(t)}
\label{eq:SGl_V}
\end{equation}
with
\begin{equation}
H_0 = -\frac{\hbar^2}{2\mn}\frac{\partial^2}{\partial z^2}+\mn gz\quad.
\label{eq:H_0}
\end{equation}
$V(t)$ is the periodic potential applied to induce the transitions. For magnetically induced transitions, $V(t) = - {\vec\mu} \cdot {\vec B}$. The eigenvectors of the solution to the Schroedinger equation with Hamiltonian $H_0$ are $\Ket{m}$.  
%
%
With transition potential, the solutions can still be written as a superposition of these eigenvectors, but with coefficients that are time-dependent. Neutrons that start out to be in the $m^{\textrm{th}}$ vertical quantum state will undergo transitions to other vertical states.

%
We use the ansatz
\begin{equation}
\Ket{\psi(t)}=\sum\limits_{m=1}^\infty a_{m}(t)\eH{-\frac{i}{\hbar}E_mt} \Ket{m}\quad.
\label{eq:Superposition}
\end{equation}
We define $\omega_{ml}=(E_l-E_m)/\hbar$. The coefficients $a_{m}(t)$ have to fulfill the coupled equations
\begin{equation}
\frac{da_{m}}{dt}=-\frac{i}{\hbar} \sum\limits_{l=1}^\infty a_{l}\eH{-i\omega_{ml}t}
\Braket{m|V(t)|l}\quad.
\label{eq:ODE_V}
\end{equation}
We note that $V(t)$ needs not only to oscillate in time, but also to contain some dependence on $z$; otherwise it would not couple different quantum states. An oscillating uniform magnetic field would not induce transitions between quantum states.

\section{Stern-Gerlach shift}
In this section we discuss a frequency shift due to the magnetic field configuration in the setups (A) and (B) described above. The oscillating magnetic field is given by a combination of a magnetic holding field $\vec B_0$ and a magnetic field that is produced by a system of parallel wires shown in Fig. \ref{fig:GRANITSetup} and which oscillates in the rest frame of the neutron. In appendix \ref{sec:MagneticFieldParallelWires}, we show how to compute the magnetic field as a function of $x$ and $z$.

In both setups (A) and (B), the neutron spin is following the magnetic field direction adiabatically. For setup (A), this has been shown in appendix \ref{sec:SetupAdiabaticity}. For setup (B), this has been shown in \cite{Pign14}. Hence, we know that $-\vec \mu_{\rm n}\cdot \vec B = +\Spm \mu_{\rm n} \cdot |\vec B|$, where $\Spm=+1$ for "spin-up" neutrons, and $\Spm=-1$ for "spin-down"-neutrons (the magnetic moment of the neutron points opposite to its spin). Through this paper, we will use $\mun=\KlB{\vec \mu_{\rm n}}$, which makes $\mun$ a positive number. We can disregard the spin state motion, as the spin stays aligned with the magnetic field, and does not affect the computation other than through the sign in $\Spm$. We can give a simplified Schroedinger equation:
\begin{equation}
H\Ket{\psi(t)}=H_0\Ket{\psi(t)}+\Spm \mun |\vec B|\Ket{\psi(t)}=i\hbar\frac{\partial}{\partial t}\Ket{\psi(t)}\quad.
\label{eq:SGlAd}
\end{equation}

A magnetic field component that oscillates, but does not depend on $z$, does not introduce transitions, as the matrix elements $\braket{m| {\Spm \mun |{\vec B}|} |l}$ vanish for $m \ne l$ . Therefore, we need to expand $|\vec B|$ at least up to the linear term in $z$:
\begin{eqnarray}
\KlB{\vec B(z,t)}&=&\KlB{\vec B(z=0,t)}+z\frac{\partial}{\partial z} \KlB{\vec B(z,t)}\nonumber \\
&=& \KlB{\vec B(z=0,t)}
    +z\frac{B_x(z=0,t)\frac{\partial B_x}{\partial z}+B_z(z=0,t)\frac{\partial B_z}{\partial z}}{\KlB{\vec B(z=0,t)}}\nonumber \\
&=& \sqrt{\Kla{\hat B_x \sin \omega t}^2+\Kla{B_0+\hat B_z \cos \omega t}^2}
    +z\frac{\Kla{-\hat B_x \sin\omega t}\Kla{-\beta_x \sin \omega t} + \Kla{B_0+\hat B_z \cos\omega t}\beta_z \cos\omega t}{\sqrt{\Kla{\hat B_x \sin \omega t}^2+\Kla{B_0+\hat B_z \cos \omega t}^2}} \quad.
\label{eq:LinearExpansionMagnitudeMagField}
\end{eqnarray}
We use this in the magnetic part of the Hamiltonian, $\Spm \mun |\vec B|$, and expand in Fourier components. We get:
\begin{equation}
\Spm \mun |\vec B|= \Spm \mun \underbrace{\Kla{\alpha_0+\alpha_1 \cos \omega t+ \alpha_2 \cos 2\omega t+ \dots}}_{\hbox{$|\vec B(0,t)|$}}+\Spm \mun \underbrace{\Kla{\beta_0+\beta_1 \cos \omega t+ \beta_2 \cos 2\omega t+ \dots}}_{\hbox{$\left.\frac{\partial}{\partial z} |\vec B(z,t)|\right|_{z=0}$}}z+\dots
\label{eq:MagFieldExpansion}
\end{equation}
Out of these, we neglect terms that oscillate with frequency $2\omega$ and above. We disregard the term $\Spm \mun \alpha_0$: This term is constant in space and time, and it just gives a constant contribution to the energy. Furthermore, we disregard the term $\Spm \mun \alpha_1 \cos \omega t$ for reasons that will become clear later. The term $\Spm \mun \beta_1 z\cos \omega t$ depends on $z$, and oscillates with $\omega$. This term is the primarily responsible one for quantum state transitions. In setup (A), in the high field limit discussed in Refs. \cite{Kre09,Pig09,Bae11a}, it reduces to $\Spm \mun \beta_z z\cos \omega t$. For a lower holding field $B_0$, it is reduced, as shown in eqs. \ref{eq:Def_Beta_i}. In our discussion, we use $B_0=1.5\Unit{mT}$, and the field values given in appendix \ref{sec:MagneticFieldParallelWires}.

The term $\Spm \mun \beta_0 z$ is not oscillating, and can be absorbed into a redefined unperturbed Hamiltonian $H_{0,\pm}$  in \eqref{eq:SGlAd} through a redefinition of $g$:
\begin{eqnarray}
H_{0,\pm} & = & -\frac{\hbar^2}{2\mn}\frac{\partial^2}{\partial x^2}+\mn g_\pm z \\
\hbox{with }g_\pm &=& g+\Spm\frac{\mun}{\mn} \beta_0 
                  \sim g\Kla{1\pm 0.125}  \quad .                   
\end{eqnarray}
Consequences of the redefined gravitational acceleration are new basis vectors ($\ket{m\pm}$) to new energies $E_{m,\pm}=E_m(g_\pm/g)^{2/3}$. Numbers given here and in the remainder of this section are for setup (A), where the transition of interest is $3\to 1$. The resonance frequency $\omega_{13,\pm}:=(E_{3,\pm}-E_{1,\pm})/\hbar$ is given by
\begin{equation}
\omega_{13,\pm} = \omega_{13}\Kla{1+\Spm\frac{\mun}{\mn g} \beta_0}^{2/3}\quad.
\end{equation}
We note that there is a shift in the resonance frequency whose sign and relative magnitude does not depend on the quantum states involved in the transition: 
\begin{equation}
\Delta \omega_{13} = \omega_{13,\pm}-\omega_{13} \sim \left\{
\begin{array}{ll}
+238\Unit{rad/s} & \text{ for "spin-up"}\quad . \\
-249\Unit{rad/s} & \text{ for "spin-down"}\quad . 
\end{array}
\right. 
\label{eq:SternGerlachShift}
\end{equation}
For an unpolarized neutron beam, in setup (A) and (B) one will encounter two distinct peaks for each transition $m\leftrightarrow l$, corresponding to the resonance frequencies for "spin-up" and "spin-down" neutrons, respectively. A precise measurement of the energies of the gravitationally bound quantum states from a single peak can be obtained if the magnetic field is known well enough to determine the size of the frequency shift. However, better than relying on the precision of a magnetic field map, is to take the following average that is not sensitive to the frequency shift:
\begin{equation}
\omega_{13}=\Kla{\frac{\omega_{13,+}^{3/2}+\omega_{13,-}^{3/2}}{2}}^{2/3}
\end{equation}
In this way, the precision of the measurement is not limited by this effect.

The cause of the frequency shift is the combination of magnetic holding field and rotating magnetic field that causes a magnetic force on the neutron spin which does not cancel when averaged over the rotation period of the magnetic field. We want to call this frequency shift the Stern-Gerlach shift. In first order, it is proportional to the magnetic moment and a magnetic field gradient, and it changes sign with the spin orientation, as seen from  
\begin{eqnarray}
\frac{\Delta \omega_{13}}{\omega_{13}}&=& \frac{2}{3} \Spm\frac{\mun}{\mn g} \beta_0
                   -\frac{1}{9} \Kla{\frac{\mun}{\mn g} \beta_0}^2
                   +\Spm\frac{4}{3}\Kla{\frac{\mun}{3\mn g} \beta_0}^3 + \dots \nonumber \\
                &=&\pm\frac{\mun\beta_x{\hat B_x}}{3\mn g B_0}
                   -\Kla{\frac{\mun\beta_x{\hat B_x}}{6\mn g B_0}}^2
                   \pm\frac{-3\mun\beta_x {\hat B_x}^3
                            +2\mun\beta_z {\hat B_x}^2{\hat B_z}
                            +2\mun\beta_x {\hat B_x}{\hat B_z}^2}{24\mn g {B_0}^3}  
                   \nonumber \\
& &  \pm\frac{1}{6}\Kla{\frac{\mun\beta_x {\hat B_x}}{3\mn g B_0}}^3+\dots
=\pm 0.0784 -0.0015 \pm 0.0044 \pm 0.0001 -\dots \quad .
\end{eqnarray}

If we use $H_{0,\pm}$ in the Schrödinger equation \eqref{eq:SGlAd}, we arrive at
\begin{equation}
H\Ket{\psi(t)}=H_{0,\pm}\Ket{\psi(t)}+\Spm \mun \beta_1 z\cos\omega t\Ket{\psi(t)}=i\hbar\frac{\partial}{\partial t}\Ket{\psi(t)} \quad.
\label{eq:SGMod}
\end{equation}
We find its solution with the following ansatz:
\begin{equation}
\Ket{\psi(t)}=\sum\limits_{m=1}^\infty a_m^\pm(t)\eH{-\frac{i}{\hbar}E_{m,\pm}t} \Ket{m\pm} \quad . 
\end{equation}
The coefficients $a_m^\pm(t)$ need to fulfill the coupled equations 
\begin{equation}
\frac{da_m^\pm}{dt}=-\frac{i}{\hbar} \sum\limits_{l=1}^\infty a_l^\pm\eH{-i\omega_{ml,\pm}t}\Braket{m\pm|\Spm \mun \beta_1 z|l\pm}\cos\omega t \quad.
\end{equation}
We retain only the first and third coefficients, as transitions between states are only possible close to a resonance, and the resonances are distinct. We will validate this approximation later.
\begin{eqnarray}
\frac{da_3^\pm}{dt}&=&-\frac{i}{\hbar} a_1^\pm\eH{i\omega_{13,\pm}t}\Braket{3\pm|\Spm \mun \beta_1 z|1\pm}\cos\omega t
-\frac{i}{\hbar} a_3^\pm\Braket{3\pm|\Spm \mun \beta_1 z|3\pm}\cos\omega t \nonumber\\
\frac{da_1^\pm}{dt}&=&-\frac{i}{\hbar} a_3^\pm\eH{-i\omega_{13,\pm}t}\Braket{1\pm|\Spm \mun \beta_1 z|3\pm}\cos\omega t
-\frac{i}{\hbar} a_1^\pm\Braket{1\pm|\Spm \mun \beta_1 z|1\pm}\cos\omega t 
\label{eq:ODE_Mod_wFast}
\end{eqnarray}
We call the terms that couple a basis state with itself the self-coupling terms. The coefficients are:
\begin{eqnarray}
\Omega_{1\pm} &:=&\frac{1}{\hbar}\Braket{1\pm|\Spm \mun \beta_1 z|1\pm}
\sim \left\{
\begin{array}{ll}
+394\Unit{rad/s} & \text{ for "spin-up"}\\
-428\Unit{rad/s} & \text{ for "spin-down"}
\end{array} \right. \\
\Omega_{3\pm} &:=&\frac{1}{\hbar}\Braket{3\pm|\Spm \mun \beta_1 z|3\pm}
\sim \left\{
\begin{array}{ll}
+930\Unit{rad/s} & \text{ for "spin-up"}\\
-1012\Unit{rad/s} & \text{ for "spin-down"}
\end{array} \right.
\end{eqnarray}
For this computation, we have used $\braket{m\pm|z|m\pm}=\braket{m|z|m}(g_\pm/g)^{-1/3}$, and eq. \eqref{eq:Matrix_z}. The coefficient of the first terms is a Rabi-Frequency, modified by the Stern-Gerlach shift:
\begin{equation}
\Omega_{31,\pm}=\frac{1}{\hbar}\Braket{1\pm|\Spm \mun\beta_1 z|3\pm}\sim \left\{
\begin{array}{ll}
-50\Unit{rad/s} & \text{ for "spin-up"}\\
54\Unit{rad/s} & \text{ for "spin-down"}
\end{array}
\right.
\end{equation}
Note that the Rabi frequency limit for high magnetic holding field, $\Omega_{31}=\Spm \mun \beta_z \Braket{1|z|3}=\mp 63.7\Unit{rad/s}$, differs by the use of unmodified eigenfunctions, and by the reduction of $\beta_1$ to $\beta_z$.

To be able to remove the self-coupling terms, we substitute $a_m^\pm\to a_m^\pm\exp(i\Omega_{m\pm}\sin\omega t/\omega)$. The new $a_m^\pm$ fulfill
\begin{eqnarray}
\frac{da_3^\pm}{dt}&=&-i \Omega_{31,\pm} a_1^\pm\eH{i\omega_{13,\pm}t}\cos\omega t
\cdot\eH{i\frac{\Omega_{3\pm}-\Omega_{1\pm}}{\omega}\sin{\omega t}} \quad ,\nonumber\\
\frac{da_1^\pm}{dt}&=&-i \Omega_{31,\pm} a_3^\pm\eH{-i\omega_{13,\pm}t}\cos\omega t
\cdot\eH{i\frac{\Omega_{1\pm}-\Omega_{3\pm}}{\omega}\sin{\omega t}} \quad . 
\end{eqnarray}
We note that any additional term in the interaction that gives an equal contribution to $\Omega_{1\pm}$ and $\Omega_{3\pm}$ exactly vanishes. That allowed us to disregard $\Spm \mun \alpha_1 \cos \omega t$ in eq. \eqref{eq:MagFieldExpansion}: This term adds to the self-coupling, but its contribution is independent of the state and does change nothing but a global phase. 
Our next step is to average over fast-oscillating terms, that is, the terms that oscillate with $\omega$ or faster. This is called the "rotating wave approximation" in atomic physics:  Above, we have two kinds of these terms. First, we replace the last factor on the right side of the last equation system with its average value:
\begin{equation}
\overline{\eH{i\frac{\Omega_{3\pm}-\Omega_{1\pm}}{\omega}\sin{\omega t}}},\overline{\eH{i\frac{\Omega_{1\pm}-\Omega_{3\pm}}{\omega}\sin{\omega t}}}\sim 1-\frac{\Kla{\Omega_{3\pm}-\Omega_{1\pm}}^2}{4\omega^2}
\label{eq:SelfCouplingAverage}
\end{equation}
In Ref. \cite{Shir63}, the quality of this approximation has been studied. The result is that to first order in the size of the neglected term, there is no frequency shift of the observed resonance. Their consequence is only that the solutions show small oscillations with frequency $\omega$
, which we also see in our numerical simulation of eqs. \eqref{eq:ODE_wSPIN}.

Close to resonance ($\omega\sim\omega_{31,\pm}$), the right side of eq. \eqref{eq:SelfCouplingAverage} evaluates to about $99\%$. This is too small to be observed. Therefore, in the rotating wave approximation, the self-coupling terms can be neglected.

The second use of the rotating wave approximation is that we write $\cos \omega t=(\exp(i\omega t)+\exp(-i\omega t))/2$ and retain only the slowly varying component in the differential equation system. We arrive at:
\begin{eqnarray}
\frac{da_3^\pm}{dt}&=&-\frac{i}{2}\Omega_{31,\pm}\eH{-i\Kla{\omega-\omega_{13,\pm}}t}a_1^\pm  \nonumber\\
\frac{da_1^\pm}{dt}&=&-\frac{i}{2}\Omega_{31,\pm}\eH{i\Kla{\omega-\omega_{13,\pm}}t}a_3^\pm  \nonumber\\
\label{eq:ODE_Mod}
\end{eqnarray}
The effect of the fast rotating components has already been studied in Ref. \cite{Blo40}. They lead to the so-called Bloch-Siegert shift in the observed resonance frequency. The Bloch-Siegert shift is too small to be observed in our experiment.

The general solution to eq. \eqref{eq:ODE_Mod} is
\begin{equation}
\Kla{\begin{array}{c} a_3^\pm(t) \\ a_1^\pm(t) \end{array}}
= \Kla{\begin{array}{cc} u_{33} & u_{31} \\ u_{13} & u_{11} \end{array}}
\cdot \Kla{\begin{array}{c} a_3^\pm(0) \\ a_1^\pm(0) \end{array}}
\quad.
\label{eq:ODE_Mod_Solution}
\end{equation}
The coefficients $u_{ml}$ are given through
\begin{eqnarray}
u_{33} &=& \Klb{\cos\Kla{\sqrt{\Kla{\omega-\omega_{13,\pm}}^2+\Omega_{31,\pm}^2} \frac{t}{2}}
            +i\frac{\omega-\omega_{13,\pm}}{\sqrt{\Kla{\omega-\omega_{13,\pm}}^2+\Omega_{31,\pm}^2}}
             \sin\Kla{\sqrt{\Kla{\omega-\omega_{13,\pm}}^2+\Omega_{31,\pm}^2} \frac{t}{2}}}
            \nonumber \\
       && \cdot\eH{-\frac{i}{2}\Kla{\omega-\omega_{13,\pm}}t} \quad , \nonumber \\
u_{31} &=& \Klb{
             -i\frac{\Omega_{31,\pm}}{\sqrt{\Kla{\omega-\omega_{13,\pm}}^2+\Omega_{31,\pm}^2}}
             \sin\Kla{\sqrt{\Kla{\omega-\omega_{13,\pm}}^2+\Omega_{31,\pm}^2} \frac{t}{2}}}
           \eH{-\frac{i}{2}\Kla{\omega-\omega_{13,\pm}}t} \quad ,\nonumber \\
u_{11} &=& \Klb{\cos\Kla{\sqrt{\Kla{\omega-\omega_{13,\pm}}^2+\Omega_{31,\pm}^2} \frac{t}{2}}
            -i\frac{\omega-\omega_{13,\pm}}{\sqrt{\Kla{\omega-\omega_{13,\pm}}^2+\Omega_{31,\pm}^2}}
             \sin\Kla{\sqrt{\Kla{\omega-\omega_{13,\pm}}^2+\Omega_{31,\pm}^2} \frac{t}{2}}}
            \nonumber \\
       && \cdot\eH{+\frac{i}{2}\Kla{\omega-\omega_{13,\pm}}t} \quad ,\nonumber \\
u_{13} &=& \Klb{
            -i\frac{\Omega_{31,\pm}}{\sqrt{\Kla{\omega-\omega_{13,\pm}}^2+\Omega_{31,\pm}^2}}
             \sin\Kla{\sqrt{\Kla{\omega-\omega_{13,\pm}}^2+\Omega_{31,\pm}^2} \frac{t}{2}}} 
            \eH{+\frac{i}{2}\Kla{\omega-\omega_{13,\pm}}t} \quad . 
\end{eqnarray}           
The evolution of the ground state population is described by a modified Rabi formula, where the modification is the use of a spin-dependent resonance frequency $\omega_{13,\pm}$ and a spin-dependent Rabi frequency $\Omega_{31,\pm}$ that are both different from their value at high magnetic holding fields. In particular, for a neutron that starts in state 3 (that is, $a_3^\pm(0)=1, a_1^\pm(0)=0$), we get:
\begin{equation}
P_{3\to 1}=\KlB{a_1^\pm(t)}^2=\frac{\sin^2\Kla{\sqrt{\Kla{\omega-\omega_{13,\pm}}^2+\Omega_{31,\pm}^2} \frac{t}{2}}}{1+\Kla{\frac{\omega-\omega_{13,\pm}}{\Omega_{31,\pm}}}^2}
\label{eq:RabiFormulaMod}
\end{equation}

For comparison with our simulations, we apply an additional correction. We note that the next term in the power series for magnetic field magnitude, eq. \eqref{eq:LinearExpansionMagnitudeMagField}, is
\begin{equation}
\frac{z^2}{2}\frac{\partial^2}{\partial z^2} \KlB{\vec B(z,t)}
= z^2\Klb{\frac{\beta_{xz}{\hat B_x}}{4B_0}+\frac{\beta_{x}^2}{4B_0}+\dots}\quad.
\end{equation}
In first order perturbation theory, the terms shown lead to an additional energy shift of each state. Using eq. \eqref{eq:Matrix_zz}, we get
\begin{eqnarray}
\Braket{m\pm|\Dmag z^2|m\pm} &=& \frac{8\Dmag}{15}z_m^2(g_\pm/g)^{-2/3} \\
\text{with } \Dmag&=& \Spm \mun \underbrace{\Klb{\frac{\beta_{xz}{\hat B_x}}{4B_0}
                                                           +\frac{\beta_{x}^2}{4B_0}}}
                     _{66\Unit{T/m$^2$}+60\Unit{T/m$^2$}} \quad .\nonumber
\end{eqnarray}
We correct the resonance frequency for that energy shift
\begin{equation}
\omega_{13,\pm}^\prime =\omega_{13,\pm}+\frac{8\mun\Dmag}{15\hbar}\Kla{z_3^2-z_1^2}(g_\pm/g)^{-2/3}\sim \left\{
\begin{array}{ll}
+9\Unit{rad/s} & \text{ for "spin-up"}\quad .\\
-11\Unit{rad/s} & \text{ for "spin-down"}\quad .
\end{array} \right. 
\label{eq:CorrectedSternGerlachShift}
\end{equation}

\begin{figure}[htb]
\includegraphics{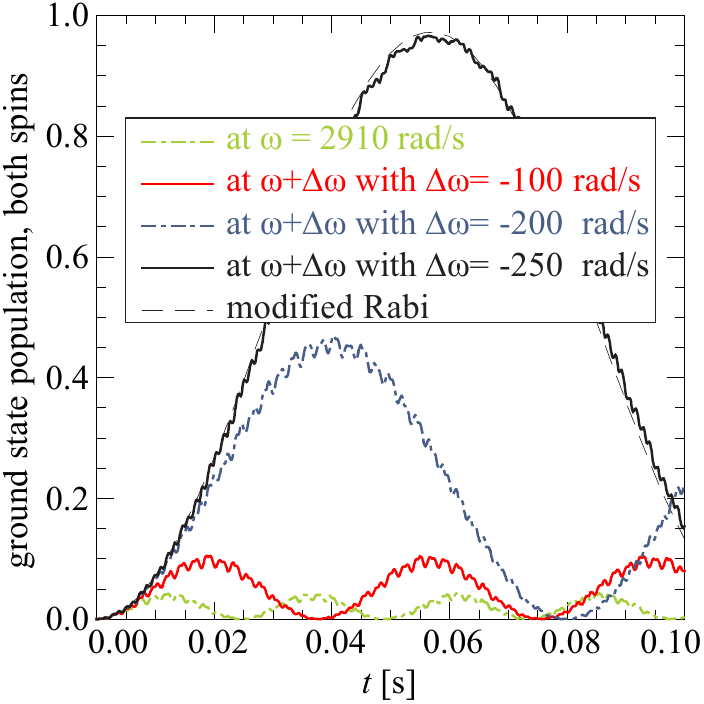}
\caption{Ground state population for a neutron that is initially in state three and with "spin-down", for different frequencies. A resonance is found for $\omega\sim 2660$\Unit{rad/s}. The line denoted "modified Rabi" is developed in this section.}
\label{fig:StateEvolution} 
\end{figure}

Fig. \ref{fig:StateEvolution} shows the result of a numerical solution of eqs. \eqref{eq:ODE_wSPIN} for neutrons that are initially "spin-down", and in state three. The simulation includes the quantum states one to five, and both spin states. If the rotation frequency $\omega$ of the magnetic field is chosen to be close to a resonance frequency, population is transfered between the quantum states at resonance. The rotation frequency is chosen to introduce transitions between states $1\leftrightarrow 3$, and the figure shows the appearance (and later disappearance) of the ground state population $\KlB{a_{1\uparrow}}^2+\KlB{a_{1\downarrow}}^2$. For the optimum frequency, the ground state population oscillated between 0 and nearly 100\%. Further away from the resonance, the ground state population stays low.
%
%
In addition, Fig. \ref{fig:StateEvolution} shows the expectation from eq. \eqref{eq:RabiFormulaMod} (dashed line). The analytic function reproduces the numerical simulation well. In the high field limit, the maximum ground state population is achieved after $t=\pi/\Omega_{31}=0.049$\Unit{s} on resonance, that is, for a resonance frequency of $\omega_{13}=2910$\Unit{rad/s}. We observe that the population change at resonance is slower for our magnetic field than it would be for a high magnetic holding field, consistent with the expectation of a Stern-Gerlach shift that we just developed.

\begin{figure}[htb]
\includegraphics{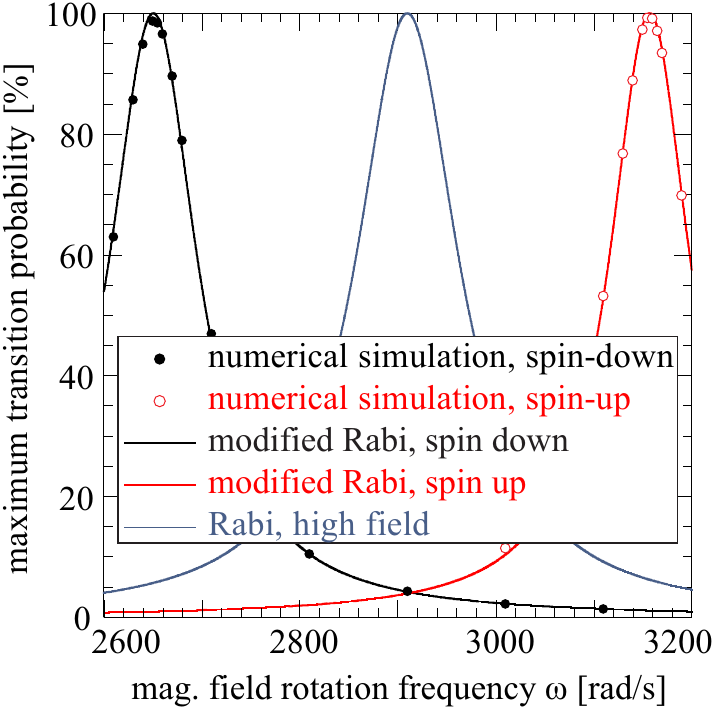}
\caption{Maximum transition probability as a function of the rotation frequency of the magnetic field $\omega$. The full (black) bullets on the left are obtained in a numerical simulation of eqs. \eqref{eq:ODE_wSPIN} for "spin-down", and the open (red) bullets are for "spin-up". The lines through the bullets give the expectation from the modified Rabi formula. For comparison, we show in the center (blue) the expectation for high magnetic field, that is, without Stern-Gerlach shift.}
\label{fig:TransitionBothSpins} 
\end{figure}

In Fig. \ref{fig:TransitionBothSpins}, we show the maximum transition probability into state 1 for neutrons that are initially in state 3, and "spin-up" (black) or "spin-down" (red). The maximum is taken over a time longer than an oscillation period. The observed Stern-Gerlach shift confirms our analytical estimate in eq. \eqref{eq:CorrectedSternGerlachShift} in sign and magnitude to about 0.1\%. An interesting outcome of this discussion is that the through-going neutrons are polarized, with opposite spin states for the two horizontal velocities classes that belong to the two peaks in Fig. \ref{fig:TransitionBothSpins}.

In addition, Fig. \ref{fig:TransitionBothSpins} shows the expectation for high magnetic holding fields ($B_0>30$\Unit{mT}) that is discussed in \cite{Kre09,Pig09,Bae11a}; in this limit the Stern-Gerlach shift is negligible, and the dynamics of the system does not depend on the spin state.

\section{Spectator state shift}

Another type of frequency shift is what we call the spectator state shift. The issue is that the rotating magnetic field couples not only the two states that are in resonance. The coupling to other energy states leads to a resonance frequency shift that is present in any measurement scheme. The spectator state shift is is formally similar to the AC stark shift in atomic physics. 

In the following, we will use the formalism of quasienergies, based on Floquet's theorem. We can write solutions of Eq. \eqref{eq:SGl_V}, in position space, in the form
\begin{equation}
\Psi(z,t)=\sum\limits_{m=1}^\infty a_m \eH{-\frac{i}{\hbar}\QE_m t} f_m(z,t) \quad .
\end{equation}
Here, $\QE_m$ are quasienergies. The theorem guarantees that $f_m(z,t)$ is periodic in time, with the same period as the periodic potential, $2\pi/\omega$. We can expand this function in a Fourier series:
\begin{equation}
\Psi(z,t)=\sum\limits_{m=1}^\infty a_{m} \eH{-\frac{i}{\hbar}\QE_mt} \sum\limits_{k=-\infty}^{\infty} \eH{-ik\omega t} \phi_m^k(z)\quad.
\label{eq:Superposition_Floquet}
\end{equation}
Using $V(z,t)=\cos{\omega t}\cdot U(z)$, the quasienergy harmonics $\phi_m^k(z)$ have to satisfy
\begin{equation}
\Kla{\QE_m+k\hbar\omega}\phi_m^k(z) = H_0 \phi_m^k(z) + \frac{1}{2}U(z)\Kla{\phi_m^{k-1}(z)+\phi_m^{k+1}(z)} \quad .
\label{eq:ODE_Floquet}
\end{equation}
This is an equation system for each positive integer $m$, and for each integer $k$. The amplitudes $a_m$ are found from the initial condition. For the most simple case, that is, for neutrons that begin in state $\ket{n_0}$, the amplitudes can be written as
\begin{equation}
a_m = \sum\limits_{k=-\infty}^{\infty} \Braket{\phi_m^k|n_0} \quad .
\label{eq:OccupationNumbers}
\end{equation} 
Thus, the transition amplitude $T_{n_0\to n_1}$ from initial state $n_0$ to final state $n_1$ after time $t$ turns out to be
\begin{equation}
T_{n_0\to n_1} = \sum\limits_{m=1}^\infty \sum\limits_{k,k^\prime=-\infty}^{\infty} 
\Braket{n_1|\phi_m^{k^\prime}} \eH{-\frac{i}{\hbar}\QE_mt-ik^\prime\omega t} \Braket{\phi_m^k|n_0} \quad .
\end{equation}
It is instructive to see how the Rabi formula can be obtained with restricting the sum to the two quasi-harmonics $n_0<n_1$. Such an approximation is justified for weak ($\KlB{\Omega_{n_0 n_1}} \ll \KlB{\omega_{n_0 n_1}}$), but resonant coupling of levels $n_0$ and $n_1$, that is with a detuning $\delta:=\omega-(E_{n_1}-E_{n_0})/\hbar$ which is small ($\KlB{\delta} \ll \KlB{\omega_{n_0 n_1}}$). 
Eq. \eqref{eq:ODE_Floquet} is approximated by
\begin{eqnarray}
\QE_{n_0}\phi_{n_0}^0(z) &=& H_0\phi_{n_0}^0(z) + \frac{1}{2}U(z)\phi_{n_0}^1(z) \quad, \nonumber \\
\Kla{\QE_{n_0}+\hbar\omega}\phi_{n_0}^1(z) &=& H_0 \phi_{n_0}^1(z) + \frac{1}{2}U(z)\phi_{n_0}^0(z) \quad .
\label{eq:ODE2_Floquet}
\end{eqnarray}
We are looking for a zeroth order solution for which $\phi_{n_0}^0(z) \approx C_0 \Braket{z|n_0}$ and $\phi_{n_0}^1(z) \approx C_1 \Braket{z|n_1}$. With this ansatz, and neglecting higher order terms, this equation system can be turned into the algebraic equations
\begin{eqnarray}
\Kla{\QE_{n_0}-E_{n_0}}C_0 &=& \frac{\hbar}{2}\Omega_{n_1 n_0}C_1 \quad ,\nonumber \\
\Kla{\QE_{n_0}+\hbar\omega-E_{n_1}}C_1 &=& \frac{\hbar}{2}\Omega_{n_0 n_1}C_0 \quad .
\end{eqnarray}
As defined previously, $\hbar\Omega_{n_1 n_0} = \braket{n_0|U|n_1}$. This system of equations has two solutions for quasienergy $\QE_{n_0}$ and coefficients $C_0$ and $C_1$:
\begin{eqnarray}
\frac{\QE_{n_0}^\pm}{\hbar} &=& \frac{E_{n_0}}{\hbar}-\frac{\delta}{2} \pm \frac{1}{2}\sqrt{\delta^2+\KlB{\Omega_{n_1 n_0}}^2} \label{eq:QE_ODE2_Floquet}\\
\textrm{and } \frac{C_1^\pm}{C_0^\pm}&=&\frac{\Omega_{n_1 n_0}}{\delta\pm\sqrt{\delta^2+\KlB{\Omega_{n_1 n_0}}^2}} \quad .
\end{eqnarray}
We determine the coefficients $C_0^\pm$ through the normalization condition $\int \KlB{ \phi_{n_0}^0(z) +\eH{-i\omega t} \phi_{n_0}^1(z)}^2 dz = 1$:
\begin{equation}
C_0^\pm=\frac{\delta\pm\sqrt{\delta^2+\KlB{\Omega_{n_1 n_0}}^2}}{\sqrt{\Kla{\delta\pm\sqrt{\delta^2+\KlB{\Omega_{n_1 n_0}}^2}}^2+\KlB{\Omega_{n_1 n_0}}^2}}
=\sqrt{\frac{1}{2}\Kla{1\pm\frac{\delta}{\sqrt{\delta^2+\KlB{\Omega_{n_1 n_0}}^2}}}}
\end{equation}
With Eq. \eqref{eq:OccupationNumbers}, we arrive at
\begin{eqnarray}
\Ket{\Psi(z,t)} &=& C_0^+\eH{-\frac{i}{\hbar}\QE_{n_0}^+t}\Kla{C_0^+\Ket{n_0}+\eH{-i\omega t}C_1^+\Ket{n_1}}
+ C_0^-\eH{-\frac{i}{\hbar}\QE_{n_0}^-t}\Kla{C_0^-\Ket{n_0}+\eH{-i\omega t}C_1^-\Ket{n_1}} \nonumber \\
&=& \Klb{\cos\Kla{\sqrt{\delta^2+\KlB{\Omega_{n_1 n_0}}^2}\frac{t}{2}}-i\frac{\delta}{\sqrt{\delta^2+\KlB{\Omega_{n_1 n_0}}^2}}\sin\Kla{\sqrt{\delta^2+\KlB{\Omega_{n_1 n_0}}^2}\frac{t}{2}}}
\eH{-i\Kla{\frac{E_{n_0}}{\hbar}-\frac{\delta}{2}}t}\ket{n_0} \nonumber \\
&& -i\frac{\Omega_{n_1 n_0}}{\sqrt{\delta^2+\KlB{\Omega_{n_1 n_0}}^2}}
\sin\Kla{\sqrt{\delta^2+\KlB{\Omega_{n_1 n_0}}^2}\frac{t}{2}}
\eH{-i\Kla{\frac{E_{n_1}}{\hbar}+\frac{\delta}{2}}t}\ket{n_1} \quad.
\label{eq:Solution_ODE2_Floquet}
\end{eqnarray}
We find the complete time-dependent solution to be the same as given in Eq. \eqref{eq:ODE_Mod_Solution}. If we neglect spectator states, resonance is achieved if the frequency of the perturbation potential $\omega$ coincides with the energy difference $\omega_{n_0 n_1}$. We will now show that taking into account additional harmonics in the quasienergy equation system \eqref{eq:ODE_Floquet} results in a shift of the resonance line in second order in $\Omega_{n_1 n_0}/\omega_{n_0 n_1}$.
For this, we need the first order correction to the wave function. We can compute this correction using Eq. \eqref{eq:ODE2_Floquet}. Together with the zeroth order, we obtain:
\begin{eqnarray}
\phi_{n_0}^0(z) & \approx & C_0\Braket{z|n_0}+\sum\limits_{k\ne n_0} C_{0,k} \Braket{z|k} \quad, \\
\phi_{n_0}^1(z) & \approx & C_1\Braket{z|n_1}+\sum\limits_{k\ne n_1} C_{1,k} \Braket{z|k}
\label{eq:FloqWaveFunction_2ndOrder}
\end{eqnarray}
with
\begin{eqnarray}
C_{0,k} &=& \frac{1}{2} \frac {\braket{k|U|n_1}}{\QE_{n_0}-E_k}C_1 
\approx \frac{\hbar}{2} \frac {\Omega_{n_1 k}}{E_{n_0}-E_k}C_1 \quad, \nonumber \\
C_{1,k} &=& \frac{1}{2} \frac {\braket{k|U|n_0}}{\QE_{n_0}+\hbar\omega-E_k}C_0
\approx \frac{\hbar}{2} \frac {\Omega_{n_0 k}}{E_{n_1}-E_k}C_0 \quad.
\label{eq:C1k_def}
\end{eqnarray}
Again we simplify equation system \eqref{eq:ODE_Floquet}, but we will keep two more harmonics:
\begin{eqnarray}
\Kla{\QE_{n_0}-\hbar\omega}\phi_{n_0}^{-1}(z) &=& H_0 \phi_{n_0}^{-1}(z) + \frac{1}{2}U(z)\phi_{n_0}^{0}(z) \quad, \nonumber \\
\QE_{n_0}\phi_{n_0}^0(z) &=& H_0\phi_{n_0}^0(z) + \frac{1}{2}U(z)\phi_{n_0}^{-1}(z) 
+ \frac{1}{2}U(z)\phi_{n_0}^1(z) \quad, \nonumber \\
\Kla{\QE_{n_0}+\hbar\omega}\phi_{n_0}^1(z) &=& H_0 \phi_{n_0}^1(z)
+ \frac{1}{2}U(z)\phi_{n_0}^0(z)+ \frac{1}{2}U(z)\phi_{n_0}^{2}(z) \quad, \nonumber \\
\Kla{\QE_{n_0}+2\hbar\omega}\phi_{n_0}^2(z) &=& H_0 \phi_{n_0}^2(z)
+ \frac{1}{2}U(z)\phi_{n_0}^1(z)
\label{eq:ODE4_Floquet}
\end{eqnarray} 
We can turn this equation system into two coupled equations using the Green's function approach. Greene's operator ${\hat G}(\epsilon)$ is
\begin{equation}
{\hat G}(\epsilon)=\sum\limits_k \frac{\ket{k}\bra{k}}{\epsilon-E_k} \quad.
\label{eq:GreenesOperator}
\end{equation}
Using Greene's operator, we write the two new harmonics as
\begin{eqnarray}
\phi_{n_0}^{-1}(z) &=& \frac{1}{2}{\hat G}(\QE_{n_0}-\hbar\omega)U(z')\phi_{n_0}^0(z') \nonumber \\
\textrm{and }\phi_{n_0}^2(z) &=& \frac{1}{2}{\hat G}(\QE_{n_0}+2\hbar\omega)U(z')\phi_{n_0}^1(z') \quad .
\end{eqnarray}
We obtain
\begin{eqnarray}
\QE_{n_0}\phi_{n_0}^0(z) &=& \Kla{H_0+\frac{1}{4}U(z){\hat G}(\QE_{n_0}-\hbar\omega)U(z')}\phi_{n_0}^0(z) + \frac{1}{2}U(z)\phi_{n_0}^1(z) \nonumber \\
\textrm{and }\Kla{\QE_{n_0}+\hbar\omega}\phi_{n_0}^1(z) &=& \Kla{H_0+\frac{1}{4}U(z){\hat G}(\QE_{n_0}+2\hbar\omega)U(z')} \phi_{n_0}^1(z)
+ \frac{1}{2}U(z)\phi_{n_0}^0(z) \quad .
\end{eqnarray} 
This equation system is similar to \eqref{eq:ODE2_Floquet}, with the only difference that we have an additional effective potential of the type ${\hat V}=(1/4)U{\hat G}U$. This equation system includes the second order corrections due to spectator states. We need to find the solution with all terms of second order in $\Omega_{n_0 n_1}/\omega_{n_0 n_1}$.

Using Eq. \eqref{eq:FloqWaveFunction_2ndOrder}, we obtain
\begin{eqnarray}
\Kla{\QE_{n_0}-E_{n_0}-V_{11}}C_0 &=& \frac{\hbar}{2}\Omega_{n_1 n_0}C_1
+\frac{\hbar}{2}\sum\limits_{k \ne n_1} \Omega_{k n_0} C_{1,k} \quad, \nonumber \\
\Kla{\QE_{n_0}+\hbar\omega-E_{n_1}-V_{22}}C_1 &=& \frac{\hbar}{2}\Omega_{n_0 n_1}C_0
+\frac{\hbar}{2}\sum\limits_{k \ne n_0} \Omega_{k n_1} C_{0,k} \quad.
\label{eq:ODE2_Floquet_wVxx}
\end{eqnarray}
We have used the definitions
\begin{eqnarray}
V_{11} &=& \frac{1}{4}\braket{n_0|U{\hat G}(\QE_{n_0}-\hbar\omega)U|n_0}
\approx \frac{\hbar^2}{4}\sum\limits_{k=1}^\infty \frac{\KlB{\Omega_{k n_0}}^2}{2E_{n_0}-E_{n_1}-E_k} \\
\textrm{and } V_{22} &=& \frac{1}{4}\braket{n_1|U{\hat G}(\QE_{n_0}+2\hbar\omega)U|n_1}
\approx \frac{\hbar^2}{4}\sum\limits_{k=1}^\infty \frac{\KlB{\Omega_{k n_1}}^2}{2E_{n_1}-E_{n_0}-E_k} \quad .
\end{eqnarray}
Substituting equation system \eqref{eq:C1k_def} in equation system \eqref{eq:ODE2_Floquet_wVxx}, we obtain
\begin{eqnarray}
\Kla{\QE_{n_0}-E_{n_0}-V_{11}-U_{11}}C_0 &=& \frac{\hbar}{2}\Omega_{n_1 n_0}C_1 \quad, \nonumber \\
\Kla{\QE_{n_0}+\hbar\omega-E_{n_1}-V_{22}-U_{22}}C_1 &=& \frac{\hbar}{2}\Omega_{n_0 n_1}C_0 \quad.
\end{eqnarray}
We introduced the notation
\begin{eqnarray}
U_{11} &=& \frac{1}{4}\sum\limits_{k\ne n_1} \frac{\KlB{\braket{n_0|U|k}}^2}{\QE_{n_0}+\hbar\omega-E_k}
\approx \frac{\hbar}{4}\sum\limits_{k\ne n_1} \frac{\KlB{\Omega_{k n_0}}^2}{E_{n_1}-E_k} \quad , \nonumber \\
U_{22} &=& \frac{1}{4}\sum\limits_{k\ne n_0} \frac{\KlB{\braket{n_1|U|k}}^2}{\QE_{n_0}-E_k}
\approx \frac{\hbar}{4}\sum\limits_{k\ne n_0} \frac{\KlB{\Omega_{k n_1}}^2}{E_{n_0}-E_k} \quad .
\end{eqnarray}
The solution for the quasienergies becomes:
\begin{equation}
\frac{\QE_{n_0}^\pm}{\hbar} = \frac{E_{n_0}}{\hbar}-\frac{\delta+\Kla{V_{11}-V_{22}+U_{11}-U_{22}}/\hbar}{2} \pm \frac{1}{2}\sqrt{\Kla{\delta+\Kla{V_{11}-V_{22}+U_{11}-U_{22}}/\hbar}^2+\KlB{\Omega_{n_1 n_0}}^2} \quad .
\end{equation}
We can compute the quasienergies $\QE_{n_0}^\pm$, and the low-frequency part of the wave function $\ket{\Psi(z,t)}$, by substituting $\delta$ with $\delta+\Kla{V_{11}-V_{22}+U_{11}-U_{22}}/\hbar$ in Eqs. \eqref{eq:QE_ODE2_Floquet} and \eqref{eq:Solution_ODE2_Floquet}. The maximum probability for disappearance from initial state $n_0$ is achieved for a resonance frequency that is larger than the energy difference of the two states in resonance by $\delta_{\rm max}=-\Kla{V_{11}-V_{22}+U_{11}-U_{22}}/\hbar$:
\begin{equation}
\delta_{\rm max}=\frac{\hbar}{4}\Kla{\sum\limits_{k=1}^\infty \frac{\KlB{\Omega_{k n_1}}^2}{2E_{n_1}-E_{n_0}-E_k}
+\sum\limits_{k\ne n_0} \frac{\KlB{\Omega_{k n_1}}^2}{E_{n_0}-E_k}
-\sum\limits_{k=1}^\infty \frac{\KlB{\Omega_{k n_0}}^2}{2E_{n_0}-E_{n_1}-E_k}
-\sum\limits_{k\ne n_1} \frac{\KlB{\Omega_{k n_0}}^2}{E_{n_1}-E_k}} \quad .
\label{eq:SpectatorShift}
\end{equation}
We note that the maximum probability for appearance into final state $n_1$ can be slightly different. If we had only the two states $n_0$ and $n_1$, we would find $\delta_{\rm max}=(\hbar/4)\KlB{\Omega_{n_0 n_1}}^2/(E_{n_1}-E_{n_0})$, which, in this approximation, corresponds to the Bloch-Siegert Shift \cite{Blo40}.

We now discuss the spectator state shift in setup (A) and setup (B) in the high magnetic holding field limit, that is, without taking into account the Stern-Gerlach shift introduced in the previous section. It can be re-introduced simply by treating each spin state separately; for each spin state, one would take the effective values for gravitational quantum state energies and Rabi frequencies. Table \ref{tab:DisappMagTransitions} compares the high-field limit of the spectator state shift for different transitions from Eq. \eqref{eq:SpectatorShift} with the frequency shift from a numerical simulation using many coupled states. For each transitions, the strength of the periodic potential was chosen so that the time for maximum transition $\pi/\Omega_{n_1 n_0}=0.1$\Unit{s}. The sums are converging only slowly, which means, that even quantum states with an energy considerably higher than $E_{n_1}+\hbar\omega$ contribute to the result. The results of the analytical expression \eqref{eq:SpectatorShift} and the numerical simulation are in reasonable agreement. The agreement could be improved by using an iterative procedure, as we have used unperturbed values of energy in the Green's operator. We have chosen a strength of the periodic potential that is weaker than in present experimental setups. This improves the quality of our approximations, but underestimates the size of the spectator state shift in those setups. The spectator state shift grows with the squared amplitude of the periodic potential. For example, in setup (A) with a length of 24\Unit{cm} and a periodic potential strength chosen for maximum transition probability for the $1\to 3$ transition, the frequency shifts are a factor of 4 larger. If the length was only 10\Unit{cm}, and again the interaction strength is chosen for maximum transition probability, the frequency shifts are about a factor 25 larger, that is, the correction of the resonance frequency of the $1\to 3$ transitions is about 2\%. Furthermore, we note that for short setups and therefore for a large amplitude excitation, the correction terms in Eq. \eqref{eq:SpectatorShift} may not be small enough, and the formalism cannot be used. 

\begin{table}[!ht]
  \caption{Frequency shift for maximum disappearance probability from initial state $n_0$ with a periodic potential in resonance with $n_0\to n_1$.}
  \label{tab:DisappMagTransitions}
\begin{center}
\begin{tabular}[c]{|c|c|c|c|}
\hline 
Transition & Unperturbed transition frequency $(E_{n_1}-E_{n_0})/2\pi\hbar$ & $\delta_{\rm max} /2\pi$ from Eq. \eqref{eq:SpectatorShift} & $\delta_{\rm max} /2\pi$ from simulation \\ 
\hline 
$1\to 2$ & 254.564\Unit{Hz} & 0.28\Unit{Hz} & 0.24\Unit{Hz} \\ 
$1\to 3$ & 462.979\Unit{Hz} & -0.03\Unit{Hz} & -0.04\Unit{Hz} \\ 
$1\to 4$ & 647.177\Unit{Hz} & 0.16\Unit{Hz} & 0.17\Unit{Hz} \\ 
$1\to 5$ & 815.557\Unit{Hz} & -0.29\Unit{Hz} & -0.29\Unit{Hz} \\ 
$1\to 6$ & 972.459\Unit{Hz} & -0.018\Unit{Hz} & -0.019\Unit{Hz} \\ 
$1\to 7$ & 1120.491\Unit{Hz} & 0.45\Unit{Hz} & 0.4\Unit{Hz} \\ 
\hline 
\end{tabular}
\end{center}
\end{table}

The periodic potential in setup (C) appears to be different, as was shown in Refs. \cite{Abe10,Mey02}; we have $V_{\rm vibr} = {\hat a}\mn g \sin\omega t+ i{\hat a}\hbar\omega\cos \omega t \frac{\partial}{\partial z}$ with a vibration amplitude $\hat a$. The first summand does not depend on position, and like a uniform magnetic field periodic in time discussed above Eq. \eqref{eq:SelfCouplingAverage}, it does not have a visible effect. The second summand can be discussed with the method described above, using $U_{\rm vibr}=i{\hat a}\hbar\omega \frac{\partial}{\partial z}$. Table \ref{tab:DisappVibrations} compares the analytical result with a numerical simulation, again with a vibration strength that would cause maximum transition probability after 0.1\Unit{s}. The analytical calculation is analogues to the one presented above, and uses matrix elements of the type $\braket{k|\frac{\partial}{\partial z}|l}$ that have been derived in \cite{Vor06}.

\begin{table}[!ht]
  \caption{Frequency shift for maximum disappearance probability from disappearance probability from initial state $n_0$ with a vibration in resonance with $n_0\to n_1$.}
  \label{tab:DisappVibrations}
\begin{center}
\begin{tabular}[c]{|c|c|c|c|}
\hline 
Transition & Unperturbed transition frequency $(E_{n_1}-E_{n_0})/2\pi\hbar$ & $\delta_{\rm max}/2\pi$ from Eq. \eqref{eq:SpectatorShift} & $\delta_{\rm max} /2\pi$ from simulation \\ 
\hline 
$1\to 2$ & 254.564\Unit{Hz} & 0.14\Unit{Hz} & 0.15\Unit{Hz} \\ 
$1\to 3$ & 462.979\Unit{Hz} & -0.08\Unit{Hz} & -0.09\Unit{Hz} \\ 
$1\to 4$ & 647.177\Unit{Hz} & 0.13\Unit{Hz} & 0.12\Unit{Hz} \\ 
$1\to 5$ & 815.557\Unit{Hz} & -0.31\Unit{Hz} & -0.29\Unit{Hz} \\ 
$1\to 6$ & 972.459\Unit{Hz} & -0.04\Unit{Hz} & -0.03\Unit{Hz} \\ 
$1\to 7$ & 1120.491\Unit{Hz} & 0.44\Unit{Hz} & 0.44\Unit{Hz} \\ 
\hline 
\end{tabular}
\end{center}
\end{table}

The formally similar effect in atomic physics, called AC Stark shift, has been discussed in a general way in Ref. \cite{Shir65}. In atoms, many transition matrix elements vanish due to selection rules, and the nonzero ones are very much smaller than differences between energy levels. Usually, large effects are expected only for spectator states that are separated from one of the states in the resonant transition by an amount that is very close to $\hbar\omega$. In gravitational resonance spectroscopy, many spectator states contribute to the shift, and the largest contribution is often given from a spectator state that is close to one of the resonant states, although it is not close to be in resonance.

\section{Interference shift}
\label{sec:FullDescription}

This section discusses setups (A) and (B), where the initial state is prepared by a step. The assumption that the initial quantum state of the incoming neutrons is state 3 is an idealization. We can approximate this situation by starting with incoming neutrons to be in the ground state before the step that separates region (1.) and (2.). The wave function of the incoming neutrons in region (1.) is $\psi_{\rm in}(z)$. Here, we have $\psi_{\rm in}(z)\propto \AiryAi((z-z_1-\zstep)/z_0)$ for $z>\zstep$, where $\zstep$ is the height of the step. Again, the one-dimensional description is justified for step heights in the order of tens of micrometers, as the energy differences between all relevant quantum states are so small that the change in the horizontal velocity of the neutron during a state transition can be neglected.  If the step is taken to be ideally sharp, we can use the sudden approximation to compute the coefficients in eq. \eqref{eq:Superposition} just after the step. at the beginning of region (2.):
\begin{equation}
a_m(\AfterStep):=a_{m\uparrow}(\AfterStep)=a_{m\downarrow}(\AfterStep)=\int\limits_\zstep^\infty \psi_{\rm in}^*(z)\psi_m(z) dz
\end{equation}

\begin{figure}[htb]
\includegraphics{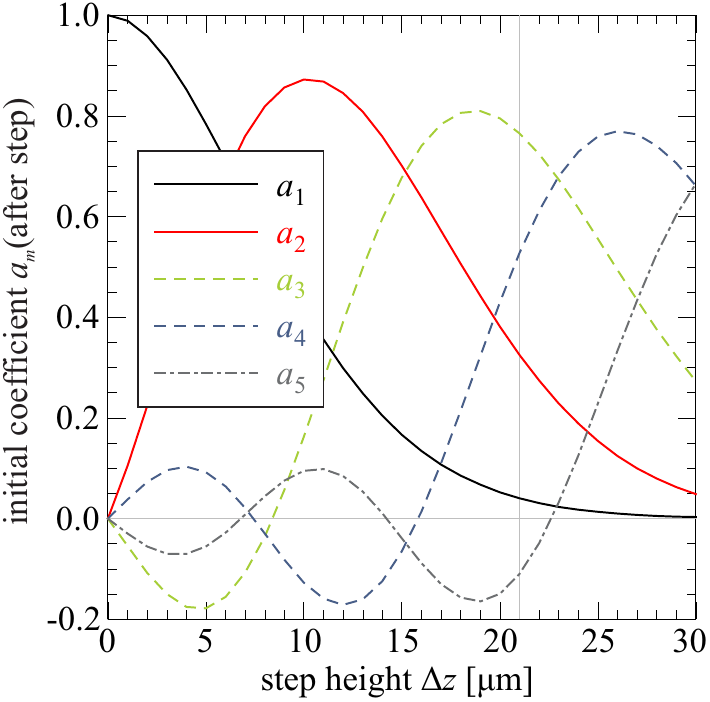}
\caption{Initial coefficients $a_m(\AfterStep)$. These coefficients are independent of the spin state, and therefore the subscript denoting the spin is suppressed.}
\label{fig:InitialCoefficients} 
\end{figure}
Fig. \ref{fig:InitialCoefficients} shows the coefficients $a_m(\AfterStep)$ for different step heights around its planned value of $\zstep=21$\UnitMu. The step height is chosen such that the coefficient $a_3(\AfterStep)$ is high compared to the other coefficients. Nevertheless, uncertainties in the determination of the step height, and the step shape, and admixtures to the incoming neutron state will cause uncertainties in the coefficients of the quantum states which are discussed in this section. For $\zstep=21$\UnitMu, we have $a_1(\AfterStep)/a_3(\AfterStep) \sim 0.05$. We have admixtures of quantum states other than 1 and 3, however, for a resonance frequency close to the $1\leftrightarrow 3$ transition, they are less important.

Neutrons traversing the distance from the step to the transition region acquire some phase that depends on the state number $m$. This phase does not change the energy levels or the resonance frequency. However, it does changes the time needed for maximum transition, and therefore it might look like a resonance peak shift, depending on the setup. In setup (A), the phase difference between state 3 and ground state, $\alpha=\arg(a_3(0))-\arg(a_1(0))$, is difficult to determine from the experiment geometry, and it might have to be treated as a fit parameter. The interference shift cannot be observed in setup (B), as here the phase of the magnetic field in the transition region changes with time. 

The filter in region (2.) will be set such that the rate of accepted neutrons $N_1(v_x)$ is proportional to $\KlB{a_1(t= L/v_x)}^2$. Using eq. \eqref{eq:ODE_Mod_Solution}, we have a count rate $N_1(v_x)$ after the filter that is given by
\begin{equation}
N_1(v_x)\propto \KlB{u_{13}}^2\KlB{a_3(0)}^2
              +2\Re\Klb{u_{13}^*u_{11}a_3^*(0)a_1(0)}
              +\KlB{u_{11}}^2\KlB{a_1(0)}^2 \quad .
\end{equation}
The elements of the matrix $u_{lm}$ are taken at $t= L/v_x$. This is the time the neutrons take to fly through the rotating field in the transition region. The first term describes how neutrons that are initially in a pure state 3 are appearing in state 1 at the end of the transition region. The last term describes how neutrons that are in a pure state 1 at the beginning of the transition region are disappearing from this state. Its contribution is reduced by $\KlB{a_1(0)/a_3(0)}^2$, and therefore small. The second term is an interference term. It disappears at resonance, and it is at maximum for $\KlB{\omega-\omega_{13\pm}^\prime}=\KlB{\Omega_{31\pm}}$. Its size is reduced by $\KlB{a_1(0)/a_3(0)}$ relative to the first term. At that level, the resonance peaks become asymmetric. This is in addition to the trivial (and small) asymmetry due to the fact that the length of the transition region $L$ can be optimized to have maximal state transfer for a certain velocity component $v_x$ in the incoming neutron beam, but not for all of them. Using eqs. \eqref{eq:ODE_Mod_Solution} and $\delta\omega(v_x)=2\pi v_x/\lambda-\omega_{13,\pm}^\prime$, we can predict the count rates at given horizontal velocity component to be
\begin{eqnarray}
N_1(v_x)&\propto&\frac{\Omega_{31,\pm}^2\KlB{a_3(0)}^2+2\Omega_{31,\pm}\delta\omega(v_x)\KlB{a_3(0)}\KlB{a_1(0)}\cos\alpha+\delta\omega^2(v_x)\KlB{a_1(0)}^2}{\delta\omega^2(v_x)+\Omega_{31,\pm}^2}\sin^2\Kla{\sqrt{\delta\omega^2(v_x)+\Omega_{31,\pm}^2}\frac{L}{2v_x}} \nonumber \\
&&+\frac{\Omega_{31,\pm}\KlB{a_3(0)}\KlB{a_1(0)}\sin\alpha}{\sqrt{\delta\omega^2(v_x)+\Omega_{31,\pm}^2}}\sin\Kla{\sqrt{\delta\omega^2(v_x)+\Omega_{31,\pm}^2}\frac{L}{v_x}} \nonumber \\
&&+\KlB{a_1(0)}^2\cos^2\Kla{\sqrt{\delta\omega^2(v_x)+\Omega_{31,\pm}^2}\frac{L}{2v_x}} \quad.
\label{eq:Nvx_model}
\end{eqnarray} 
\begin{figure}[htb]
\includegraphics{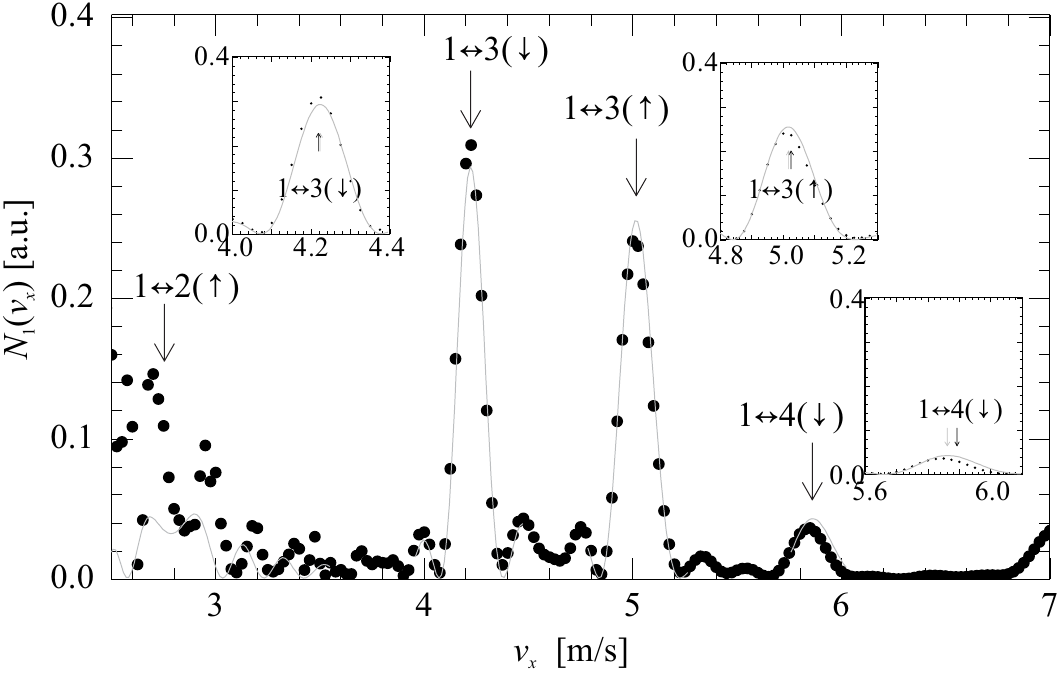}
\caption{Expected yield for different velocity components $v_x$ of the incoming neutron beam.The arrows in parenthesis denote if the resonance is for the "spin-up" ($\uparrow$) or the "spin-down" ($\downarrow$) component. The bullets are obtained in a numerical simulation of eqs. \eqref{eq:ODE_wSPIN}. The gray lines show the analytical model (eq. \eqref{eq:Nvx_model}). The arrows indicate the velocity components for which the transitions occur that are indicated above the arrows. The $v_x$ distribution has been neglected in this figure. It is expected to emphasize the higher values for $v_x$.}
\label{fig:Nvx} 
\end{figure}

Fig. \ref{fig:Nvx} shows the precision with which the model presented here allows to determine the position of the peaks. The agreement with the full simulation is well below the peak width for the peaks for which the length of the transition region is optimized for maximum transition. For the small peaks, the combination of interference effects and the non-perfect reproduction of the Rabi frequency by the model is a limitation. For "spin-up" neutrons, the $1\leftrightarrow 2$ transition is not well separated from the $2\leftrightarrow 3$, and our two-state model cannot capture the dynamics correctly. The transition $1\leftrightarrow 4$ is a case where the interference shift is large. 

The maximum of the resonance peak, $v_{x,\textrm peak}$, is shifted from its unperturbed value $\lambda\omega_{13,\pm}^\prime/2\pi$ due to the interference term. In first approximation, this shift is given by
\begin{equation}
\frac{\Delta v_{x,\textrm peak}}{v_{x,\textrm peak}} 
 =\frac{\frac{a_1(0)}{a_3(0)}\frac{\Omega_{31,\pm}}{\omega_{13,\pm}^\prime}\cos\alpha}{\Kla{1-\frac{a_1^2(0)}{a_3^2(0)}}\Kla{1+\frac{\Omega_{31,\pm}}{\omega_{13,\pm}^\prime}\frac{L\pi}{\lambda}\cot{\frac{\Omega_{31,\pm}}{\omega_{13,\pm}^\prime}\frac{L\pi}{\lambda}}}+\frac{a_1(0)}{a_3(0)}\Kla{\frac{\Omega_{31,\pm}}{\omega_{13,\pm}^\prime}\frac{L\pi}{\lambda}\Kla{\cot^2\frac{\Omega_{31,\pm}}{\omega_{13,\pm}^\prime}\frac{L\pi}{\lambda}-1}-1}\sin\alpha} \quad.
\label{eq:InterferenceShift}
\end{equation}
The interference shift is small for full transitions, where we have 
\begin{equation}
\frac{\Delta v_{x,\textrm peak}}{v_{x,\textrm peak}} 
 \sim \frac{a_1(0)}{a_3(0)} \frac{\Omega_{31,\pm}}{\omega_{13,\pm}^\prime} \cos \alpha \quad.
\end{equation}
Irrespective of the value for $\alpha$, the correction is usually below 0.1\%.  It gets large for a given resonance if the length of the transition region $L$ is not chosen such that the transition probability is at maximum, that is, if the argument of the cotangents is not close to an half-integral multiple of $\pi$. 
In Fig. \ref{fig:Nvx}, the gray arrow show the position of the maximum after taking into account the interference shift. The difference to the expectation without interference (the black arrow) is small, and most pronounced at the $1\leftrightarrow 4$ transition, where the transition probability is small since the length of the system $L$ is not optimized for that transition.  

\section{Summary}

This paper discusses systematic frequency shifts in the spectroscopic measurement of the energies of gravitationally bound quantum states of ultra-cold neutrons. The largest one, the Stern-Gerlach shift, is associated with the dynamics of the spin; it is specific to the setup presented here, in which resonant transitions between quantum states are induced with an oscillating magnetic field gradient in flow-through mode. The Stern-Gerlach shift is spin-dependent, and disappears by taking a suitable average over the spin state. The spectator state shift is present in any scheme of gravitational resonance spectroscopy. The interference shift stems from an imperfect preparation of the initial state through a step, and is averaged out in many schemes. These frequency shifts need to be discussed and taken into account in a gravitational resonance spectroscopy if precision measurements are desired.

\begin{acknowledgments}
The authors are grateful to members of the GRANIT collaboration for help
and advice.
S.B. thanks Tom Gallagher for a helpful conversation about Ref. \cite{Shir65}.
We acknowledge support from the National Science foundation NSF-0855610.
\end{acknowledgments}

\appendix
\section{Magnetic field of parallel rectangular wires}
\label{sec:MagneticFieldParallelWires}

We will now compute the magnetic field of an infinite amount of parallel rectangular wires with regular spacing and current pattern. This is the set of wires in the transition region in setups (A) and (B), except for that the real wires have finite length, are interconnected, and we have a finite amount of them. We will solve the problem in three steps: First we give an expression for an infinite number of parallel thin wires, then we extend the solution to wires with rectangular cross section, and finally we superimpose wire sets with different currents to take into account the pattern of currents we have in our setup.

The magnetic field of a single wire with current $I$, parallel to the $y$ axis, positioned at $(x_0,z_0)$, for a point $(x,z)$ outside the wire, is
\begin{equation}
B_x(x,z) = \frac{\mu_0 I}{2\pi}\frac{z-z_0}{\Kla{x-x_0}^2+\Kla{z-z_0}^2} \quad .
\end{equation}
For an infinite amount of wires at distance $\lambda$, we add the magnetic field of each one. Assuming the $n^{\textrm th}$ wire is at $x=x_0+n\lambda$, we get
\begin{equation}
B_x(x,z) = \frac{\mu_0 I}{2\pi}\sum\limits_{m=-\infty}^{\infty}\frac{z-z_0}{\Kla{x-(x_0+n\lambda)}^2+\Kla{z-z_0}^2} \quad .
\end{equation}
To understand better the $x$ and $z$ dependence of this function, let us calculate the Fourier transform with respect to the $x$ variable:
\begin{equation}
B_x(k,z) = \frac{1}{\sqrt{2\pi}}\int\limits_{-\infty}^{\infty}B_x(x,z) \eH{ikx}dx \quad .
\end{equation}
Taking into account that for $z<z_0$ (the case we are interested in), 
\begin{equation}
\int\limits_{-\infty}^{\infty}\frac{z-z_0}{\Kla{x-x_0}^2+\Kla{z-z_0}^2} \eH{ikx}dx
=-\pi\eH{ikx_0}\eH{-\KlB{k(z-z_0)}} \quad ,
\end{equation}
one obtains
\begin{equation}
B_x(k,z) = -\frac{\mu_0 I}{2\sqrt{2\pi}} \eH{-\KlB{k(z-z_0)}} \sum\limits_{n=-\infty}^{\infty} \eH{ikx_0+ikna}
= -\sqrt{2\pi} \frac{\mu_0 I}{2\lambda} \eH{-\KlB{k(z-z_0)}} \eH{ikx_0} \sum\limits_{n=-\infty}^{\infty} \delta\Kla{k-n\frac{2\pi}{\lambda}} \quad .
\end{equation}
We used here the usual expression for the Dirac comb ($T=2\pi/\lambda$):
\begin{equation}
\frac{1}{T}\sum\limits_{n=-\infty}^{\infty} \eH{i 2\pi n t/T} = \sum\limits_{n=-\infty}^{\infty} \delta\Kla{t-nT} \quad .
\end{equation}
Therefore, for the magnetic field, we obtain
\begin{equation}
B_x(x,z) = -\frac{1}{\sqrt{2\pi}}\int\limits_{-\infty}^{\infty}B_x(k,z) \eH{-ikx}dk
\end{equation}
and
\begin{equation}
B_x(x,z) = -\frac{\mu_0 I}{2\lambda}\sum\limits_{n=-\infty}^{\infty} \eH{i 2\pi n(x-x_0)/\lambda} \eH{-2\pi \KlB{n(z-z_0)}/\lambda}
= -\frac{\mu_0 I}{\lambda}\Kla{\frac{1}{2}+\sum\limits_{n=1}^{\infty} \eH{-2\pi n\KlB{z-z_0}/\lambda} \cos\frac{2\pi n(x-x_0)}{\lambda} }\quad .
\end{equation}
This field represents a series of periodic terms where higher harmonics are suppressed by a factor of $\eH{-2\pi (z-z_0)/\lambda}$:
\begin{equation}
B_x(x,z) = -\frac{\mu_0 I}{\lambda}\Kla{\frac{1}{2}+\eH{-2\pi \KlB{z-z_0}/\lambda}\cos\frac{2\pi (x-x_0)}{\lambda}+\eH{-4\pi \KlB{z-z_0}/\lambda}\cos\frac{4\pi (x-x_0)}{\lambda}+\dots} 
\end{equation}
We can rewrite this as
\begin{eqnarray}
B_x(x,z)
& = & -\frac{\mu_0 I}{\lambda}\Kla{\frac{1}{2}+\sum\limits_{n=1}^{\infty} \eH{-2\pi n\KlB{z-z_0}/\lambda} \cos\frac{2\pi (x-x_0)}{\lambda} } \nonumber \\
& = & -\frac{\mu_0 I}{2\lambda}\Kla{1+\sum\limits_{n=1}^{\infty} \eH{-2\pi n\KlB{z-z_0}/\lambda} \eH{2\pi in(x-x_0)/\lambda} +\sum\limits_{n=1}^{\infty} \eH{-2\pi n\KlB{z-z_0}/\lambda} \eH{-2\pi in(x-x_0)/\lambda} } \nonumber \\
& = & -\frac{\mu_0 I}{2\lambda}\Kla{1+\frac{\eH{-2\pi n\KlB{z-z_0}/\lambda}\eH{2\pi i(x-x_0)/\lambda}}{1-\eH{-2\pi \KlB{z-z_0}/\lambda}\eH{2\pi i(x-x_0)/\lambda}} + \frac{\eH{-2\pi \KlB{z-z_0}/\lambda}\eH{-2\pi i(x-x_0)/\lambda}}{1-\eH{-2\pi \KlB{z-z_0}/\lambda}\eH{-2\pi i(x-x_0)/\lambda}} } \nonumber \\
& = & \frac{\mu_0 I}{2\lambda} \frac{\sinh{2\pi (z-z_0)/\lambda}}{\cosh{2\pi (z-z_0)/\lambda}-\cos{ 2\pi (x-x_0)/\lambda}} \quad.
\end{eqnarray}

An analogous expression can be obtained for the $z$-component of the field:
\begin{eqnarray}
B_z(x,z)&=& -\frac{\mu_0 I}{\lambda}\Kla{\eH{-2\pi \KlB{z-z_0}/\lambda}\sin\frac{2\pi (x-x_0)}{\lambda}+\eH{-4\pi \KlB{z-z_0}/\lambda}\sin\frac{4\pi (x-x_0)}{\lambda}+\dots } \nonumber \\
&=& -\frac{\mu_0 I}{2\lambda} \frac{\sin{2\pi (x-x_0)/\lambda}}{\cosh{2\pi \KlB{z-z_0}/\lambda}-\cos{ 2\pi (x-x_0)/\lambda}}
\end{eqnarray}

The next step is to replace the infinitely thin wires at $x=x_0+n\lambda$ and $z=z_0$ by a wire with rectangular cross section of width $\Delta x=1\Unit{mm}$ and height $\Delta z=1\Unit{mm}$, centered at the same place. For $z<z_0-\Delta z/2$, we get:
\begin{eqnarray}
B_x(x,z) = -\int\limits_{x_0-\Delta x/2}^{x_0+\Delta x/2} \int\limits_{z_0-\Delta z/2}^{z_0+\Delta z/2} \frac{\mu_0 I}{\lambda\Delta x \Delta z}\left( \frac{1}{2} \right.
&+&\eH{-2\pi \KlB{z-z^\prime}/\lambda}\cos\frac{2\pi (x-x^\prime)}{\lambda} \nonumber \\
&+&\left.\eH{4\pi \KlB{z-z^\prime}/\lambda}\cos\frac{4\pi (x-x^\prime)}{\lambda}+\dots \right) dz^\prime dx^\prime \quad.
\end{eqnarray}
We evaluate this expression, to
\begin{eqnarray}
B_x(x,z) = -\frac{\mu_0 I}{\lambda}\left( \frac{1}{2} \right.
  &+& \frac{\lambda^2}{\pi^2 \Delta x \Delta z}\eH{-2\pi \KlB{z-z_0}/\lambda}\sinh\frac{\pi\Delta z}{\lambda}\cos\frac{2\pi (x-x_0)}{\lambda}\sin\frac{\pi\Delta x}{\lambda}  \nonumber \\
  &+& \left. \frac{\lambda^2}{(2\pi)^2 \Delta x \Delta z}\eH{-4\pi \KlB{z-z_0}/\lambda}\sinh\frac{2\pi\Delta z}{\lambda}\cos\frac{4\pi (x-x_0)}{\lambda}\sin\frac{2\pi\Delta x}{\lambda}
    + \dots \right) \quad .
\label{eq:BxSingleI}    
\end{eqnarray}
In the same way, we get
\begin{eqnarray}
B_z(x,z) = -\frac{\mu_0 I}{\lambda}&\left( \vphantom{\frac{1}{1}} \right.
  & \frac{\lambda^2}{\pi^2 \Delta x \Delta z}\eH{-2\pi \KlB{z-z_0}/\lambda}\sinh\frac{\pi\Delta z}{\lambda}\sin\frac{2\pi (x-x_0)}{\lambda}\sin\frac{\pi\Delta x}{\lambda} \nonumber \\
  && \left. +\frac{\lambda^2}{(2\pi)^2 \Delta x \Delta z}\eH{-4\pi \KlB{z-z_0}/\lambda}\sinh\frac{2\pi\Delta z}{\lambda}\sin\frac{4\pi (x-x_0)}{\lambda}\sin\frac{2\pi\Delta x}{\lambda}
    + \dots \right) \quad .
\end{eqnarray}

The last step is to add the magnetic field of infinite parallel wires with current $I=I_0$ at $x_0=x_\textrm{offset}=\lambda/4$ to the magnetic field of infinite parallel wires with current $I=I_0/\sqrt{2}$ at $x_0=x_\textrm{offset}+\lambda/8$, and so on up to the the magnetic field of infinite parallel wires with current $I=I_0/\sqrt{2}$ at $x_0=x_\textrm{offset}+7\lambda/8$. For the choice of currents, the terms that rotate with spatial period $\lambda/4\pi$ and $\lambda/6\pi$ cancel. Neglecting the terms that rotate even faster, we obtain:
\begin{eqnarray}  
B_x(x,z) &=& -\frac{\mu_0 I_0}{\lambda} \frac{\lambda^2}{\pi^2 \Delta x \Delta z}\eH{-2\pi \KlB{z-z_0}/\lambda}\sinh\frac{\pi\Delta z}{\lambda}\sin\frac{\pi\Delta x}{\lambda} \nonumber \\
&& \cdot \left(\cos\frac{2\pi (x-\lambda/4)}{\lambda}+\frac{1}{\sqrt{2}}\cos\frac{2\pi (x-3\lambda/8)}{\lambda}-\frac{1}{\sqrt{2}}\cos\frac{2\pi (x-5\lambda/8)}{\lambda} \right. \nonumber \\
&&\quad \left. -\cos\frac{2\pi (x-3\lambda/4)}{\lambda}-\frac{1}{\sqrt{2}}\cos\frac{2\pi (x-7\lambda/8)}{\lambda}+\frac{1}{\sqrt{2}}\cos\frac{2\pi (x-9\lambda/8)}{\lambda} \right) \nonumber \\
&=& -\frac{4\mu_0 I_0}{\lambda} \frac{\lambda^2}{\pi^2 \Delta x \Delta z}\eH{-2\pi \KlB{z-z_0}/\lambda}\sinh\frac{\pi\Delta z}{\lambda}\sin\frac{\pi\Delta x}{\lambda}
\sin\frac{2\pi x}{\lambda} \quad .
\end{eqnarray}
For the $z$ component, we get:
\begin{eqnarray}
B_z(x,z) &=& -\frac{\mu_0 I_0}{\lambda} \frac{\lambda^2}{\pi^2 \Delta x \Delta z}\eH{-2\pi \KlB{z-z_0}/\lambda}\sinh\frac{\pi\Delta z}{\lambda}\sin\frac{\pi\Delta x}{\lambda}\nonumber \\
&& \cdot \left(\sin\frac{2\pi (x-\lambda/4)}{\lambda}+\frac{1}{\sqrt{2}}\sin\frac{2\pi (x-3\lambda/8)}{\lambda}-\frac{1}{\sqrt{2}}\sin\frac{2\pi (x-5\lambda/8)}{\lambda} \right. \nonumber \\
&&\quad \left. -\sin\frac{2\pi (x-3\lambda/4)}{\lambda}-\frac{1}{\sqrt{2}}\sin\frac{2\pi (x-7\lambda/8)}{\lambda}+\frac{1}{\sqrt{2}}\sin\frac{2\pi (x-9\lambda/8)}{\lambda} \right) \nonumber \\
&=& \frac{4\mu_0 I_0}{\lambda} \frac{\lambda^2}{\pi^2 \Delta x \Delta z}\eH{-2\pi \KlB{z-z_0}/\lambda}\sinh\frac{\pi\Delta z}{\lambda}\sin\frac{\pi\Delta x}{\lambda}
\cos\frac{2\pi x}{\lambda} \quad.
\end{eqnarray}
From this, we obtain
\begin{eqnarray}
{\hat B_x} = {\hat B_z} & =& \frac{4\mu_0 I_0}{\lambda} \frac{\lambda^2}{\pi^2 \Delta x \Delta z}\eH{-2\pi \KlB{z-z_0}/\lambda}\sinh\frac{\pi\Delta z}{\lambda}\sin\frac{\pi\Delta x}{\lambda} \approx 1\Unit{mT} \quad, \\
\beta_x  = \beta_z &=& \frac{8\pi\mu_0 I_0}{\lambda^2} \frac{\lambda^2}{\pi^2 \Delta x \Delta z}\eH{-2\pi \KlB{z-z_0}/\lambda}\sinh\frac{\pi\Delta z}{\lambda}\sin\frac{\pi\Delta x}{\lambda} \approx 0.6\Unit{T/m} \quad, \\
\beta_{zz} = \beta_{xz} &=& \frac{16\pi^2\mu_0 I_0}{\lambda^3} \frac{\lambda^2}{\pi^2 \Delta x \Delta z}\eH{-2\pi \KlB{z-z_0}/\lambda}\sinh\frac{\pi\Delta z}{\lambda}\sin\frac{\pi\Delta x}{\lambda} \approx 400\Unit{T/m$^2$} \quad.
\end{eqnarray}
The numerical values for the amplitudes of magnetic field and its derivatives are given for $I_0=5\Unit{A}$, $\lambda = 1\Unit{cm}$, and for a location that is $1.5\Unit{mm}$ below the wires ($z=z_0-1.5\Unit{mm}$). In our discussions, we have used the numerical values for ${\hat B_x}, {\hat B_z}, \dots$ as independent inputs, to be able to take into account a measured magnetic field which might differ due to imperfections in the geometry.

We have shown in \cite{Bae11a} that $v_x$, the $x$ component of the neutron velocity, can be taken as constant for each neutron for the measurements planned. Therefore, we can write the magnetic field alternatively as a function of time and $z$ coordinate using the translation $x=v_x\cdot t$. The rotation frequency of the field, as seen by the neutron, is then given as $\omega=v_x/\lambda$.

For this field, the expansion coefficients used in the text in eq. \eqref{eq:MagFieldExpansion} can be further expanded in $\hat B_x/B_0$ and $\hat B_z/B_0$, and are given by:
\begin{eqnarray}
\alpha_0 &=& B_0\Kla{1+\frac{{\hat B_x}^2}{4B_0^2}
                     +\frac{-3{\hat B_x}^4+4{\hat B_x}^2{\hat B_z}^2}{64B_0^4}
                     +\dots} \sim \Kla{1.5+0.167+0.005+\dots}\Unit{mT} \quad, \nonumber \\
\alpha_1 &=& {\hat B_z}\Kla{1-\frac{{\hat B_x}^2}{8B_0^2}
                     +\frac{3{\hat B_x}^4-4{\hat B_x}^2{\hat B_z}^2}{64B_0^4}
                     +\dots} \sim \Kla{1.0-0.056-0.003+\dots}\Unit{mT} \quad, \nonumber \\
\alpha_2 &=& -\frac{{\hat B_x}^2}{4B_0}+\frac{{\hat B_x}^4}{16B_0^3}+\dots
             \sim \Kla{-0.167+0.018+\dots}\Unit{mT} \label{eq:Def_Alpha_i} \quad,\\
\beta_0 &=& \frac{\beta_x\hat B_x}{2B_0}
            +\frac{-3\beta_x{\hat B_x}^3+2\beta_x{\hat B_x}{\hat B_z}^2
                               +2\beta_z{\hat B_x}^2{\hat B_z}}{16{B_0}^3} 
            +\frac{15\beta_x{\hat B_x}^5-12\beta_z{\hat B_x}^4{\hat B_z} 
                   -24\beta_x{\hat B_x}^3{\hat B_z}^2+16\beta_z{\hat B_x}^2{\hat B_z}^3
                   +8\beta_x{\hat B_x}{\hat B_z}^4}{128{B_0}^5}\nonumber \\
         && +\dots 
         \sim \Kla{0.2+0.0111+0.0019+\dots}\Unit{T/m} \quad, \nonumber \\
%
\beta_1 &=& \beta_z
           +\frac{-\beta_z{\hat B_x}^2-2\beta_x{\hat B_x}{\hat B_z}}{8{B_0}^2}
           +\frac{3\beta_z{\hat B_x}^4+12\beta_x{\hat B_x}^3{\hat B_z}
                  -12\beta_z{\hat B_x}^2{\hat B_z}^2-8\beta_x{\hat B_x}{\hat B_z}^3}{64{B_0}^4}
             \nonumber \\      
         &\sim & \Kla{0.6-0.1-0.0093+\dots}\Unit{T/m}  \quad,     
             \nonumber \\
\beta_2 &=& \frac{-\beta_x\hat B_x}{2B_0}
            +\frac{\beta_x{\hat B_x}^3}{4{B_0}^3}
            \sim \Kla{-0.2+0.044+\dots}\Unit{T/m} \quad.
\label{eq:Def_Beta_i} 
\end{eqnarray}


\section{Equivalence of solutions to Schroedinger equation in adiabatic approximation to the original one}
\label{sec:SetupAdiabaticity}

The Schroedinger equation for a oscillating magnetic interaction potential is given by
\begin{equation}
H\Ket{\psi(t)}=H_0\Ket{\psi(z,t)}-\vec \mu_{\rm n}\cdot \vec B(t)\Ket{\psi(t)}=i\hbar\frac{\partial}{\partial t}\Ket{\psi(t)} \quad .
\label{eq:SGl}
\end{equation}

We had asserted in the main text that this equation is equivalent to the one that we used as eq. \eqref{eq:SGlAd}, that is, in adiabatic approximation. In this section we want to derive a condition for that, valid for the conditions of setup (A).

Let us define the spinor $\Ket{+}$ ($\Ket{-}$) as one that describes a neutron spin that is always aligned (anti-aligned) with the magnetic field. That means that this spinor is an eigen-vector of the magnetic interaction Hamiltonian 
\begin{equation}
-\vec \mu_{\rm n}\cdot \vec B(z,t) \Ket{\pm}= \epsilon_\pm(z,t) \Ket{\pm} \quad.
\end{equation}
Here, $z$ and $t$ play a role as parameters. For ${\vec B}(z,t)=B_x(z,t){\hat x}+B_z(z,t){\hat z}$, one easily gets $\epsilon_\pm(z,t)=\pm \mun B(z,t)$. The spinor has the components
\begin{eqnarray}
\Ket{+} &=&
  \Kla{\begin{array}{c} -\frac{B-B_z}{\sqrt{B_x^2+\Kla{B-B_z}^2}} \\ \frac{B_x}{\sqrt{B_x^2+\Kla{B-B_z}^2}} \end{array}}
  = \frac{1}{\sqrt{2}}\Kla{\begin{array}{c} -\sqrt{1-\frac{B_z}{B}} \\ \sqrt{1+\frac{B_z}{B}}\end{array}}  \quad, \\  
\Ket{-} &=&
  \Kla{\begin{array}{c} \frac{B+B_z}{\sqrt{B_x^2+\Kla{B+B_z}^2}} \\ \frac{B_x}{\sqrt{B_x^2+\Kla{B+B_z}^2}} \end{array}}
  = \frac{1}{\sqrt{2}} \Kla{\begin{array}{c} \sqrt{1+\frac{B_z}{B}} \\ \sqrt{1-\frac{B_z}{B}}\end{array}} \quad.
\end{eqnarray}
With these spinors, the general wave function is now
\begin{equation}
\Ket{\psi(t)}=\phi_+(z,t)\Ket{+}+\phi_-(z,t)\Ket{-} \quad.
\end{equation}
Taking into account that the spinors $\Ket{+}$, $\Ket{-}$ are orthogonal to each other at any moment $t$ and any position $z$, we get the corresponding system for the new wave function components
\begin{eqnarray}
H_0\phi_+(z,t)+\mun B(z,t)\phi_+(z,t) &=&i\hbar\frac{\partial}{\partial t}\phi_+(z,t)+i\hbar\phi_-(z,t)\braket{-|\frac{\partial}{\partial t}|+} \quad, \nonumber \\
H_0\phi_-(z,t)-\mun B(z,t)\phi_-(z,t) &=&i\hbar\frac{\partial}{\partial t}\phi_-(z,t)-i\hbar\braket{+|\frac{\partial}{\partial t}|-} \quad.
\end{eqnarray}
Decoupling of amplitudes that corresponding to different spin projections on the field axis (the adiabatic approximation) takes place if we neglect the last term in each equation, that is, their coupling. This approximation is valid if
\begin{equation}
\KlB{\hbar \braket{-|\frac{\partial}{\partial t}|+}} \ll \KlB{\mun B} \quad.
\label{eq:DecouplingValid}
\end{equation}
The spinors $\Ket{+}$, $\Ket{-}$ are periodic functions of $t$, and their periodicity is chosen to match the transition frequency, e.g. $\omega_{13,\pm}$:
\begin{equation}
\KlB{\hbar \braket{-|\frac{\partial}{\partial t}|+}}\sim\KlB{\frac{1}{2\alpha_0}\frac{\partial B_z}{\partial t}} \le \frac{{\hat B_z}\pi v_x}{\alpha_0\lambda}
\end{equation}
Here, $\alpha_0$ is the average value of the magnetic field magnitude $\KlB{B}$, as defined in Eq. \eqref{eq:MagFieldExpansion}. Eq. \eqref{eq:DecouplingValid} holds if the frequency with which the magnetic field changes is small compared to the Larmor frequency $2\mun \alpha_0/\hbar$:
\begin{equation}
\frac{{\hat B_z}\pi v_x}{\alpha_0\lambda} \ll \frac{2\mun \alpha_0}{\hbar}
\end{equation}
For this condition to hold for all neutron velocities up to $v_x=7$\Unit{m/s}, we need $\alpha_0\gg \sqrt({{\hat B_z}\hbar\pi v_x}/{2\lambda\mun})$. This is verified in a numerical simulation of the full Schroedinger equation \eqref{eq:SGl} including both spin states, demanding $B_0 \ge 1.5$\Unit{mT}. 

\section{Numerical simulation of transition region in setup (A):}
We have developed a full numerical simulation of the Schroedinger equation with spin \eqref{eq:SGl}. We use the ansatz:
\begin{equation}
\Ket{\psi(t)}=\sum\limits_{m=1}^\infty a_{m\uparrow}(t)\eH{-\frac{i}{\hbar}E_mt} \Ket{m\uparrow}+a_{m\downarrow}(t)\eH{-\frac{i}{\hbar}E_mt} \Ket{m\downarrow}\quad.
\end{equation}
Here, $\ket{m\uparrow}$ is the $m^{\text{th}}$ solution of eq. \eqref{eq:SGl} without magnetic field to a spin pointing into the +z direction ("spin-up") to energy $E_m$, and $\ket{m\downarrow}$ is the $m^{\text{th}}$ solution of eq. \eqref{eq:SGl} without magnetic field to a spin pointing into the -z direction ("spin-down"). We define $\omega_{ml}=(E_l-E_m)/\hbar$. The coefficients $a_{m\uparrow}(t)$ and $a_{m\downarrow}(t)$ have to fulfill the coupled equations
\begin{eqnarray}
\frac{da_{m\uparrow}}{dt}&=&-\frac{i}{\hbar} \sum\limits_{l=1}^\infty a_{l\uparrow}\eH{-i\omega_{ml}t}
\Braket{m\uparrow|-\vec \mu_{\rm n}\cdot \vec B|l\uparrow}
-\frac{i}{\hbar} \sum\limits_{l=1}^\infty a_{l\downarrow}\eH{-i\omega_{ml}t}
\Braket{m\uparrow|-\vec \mu_{\rm n}\cdot \vec B|l\downarrow} \quad, \nonumber\\
\frac{da_{m\downarrow}}{dt}&=&-\frac{i}{\hbar} \sum\limits_{l=1}^\infty a_{l\uparrow}\eH{-i\omega_{ml}t}
\Braket{m\downarrow|-\vec \mu_{\rm n}\cdot \vec B|l\downarrow}
-\frac{i}{\hbar} \sum\limits_{l=1}^\infty a_{l\downarrow}\eH{-i\omega_{ml}t}
\Braket{m\downarrow|-\vec \mu_{\rm n}\cdot \vec B|l\uparrow} \quad.
\label{eq:ODE_wSPIN}
\end{eqnarray}

For the computation of the matrix elements, we need to write $-\vec \mu_{\rm n}\cdot \vec B = \mun\vec \sigma\cdot \vec B$, where $\sigma_x$, $\sigma_y$, and $\sigma_z$ are the Pauli matrices. For the piece of the magnetic field linear or quadratic in $z$, we have used:
\begin{eqnarray}
\braket{m\uparrow|z|l\uparrow}=\braket{m\downarrow|z|l\downarrow}&=&\left\{
\begin{array}{ll}
(2/3) z_m & \text{ for } m=l\\
2(-1)^{m-l+1}z_0^3/(z_m-z_l)^2 & \text{ otherwise}
\end{array} \right. 
\label{eq:Matrix_z} \\
\textrm{and } \braket{m\uparrow|z^2|l\uparrow}=\braket{m\downarrow|z^2|l\downarrow}&=&\left\{
\begin{array}{ll}
(8/15) z_m^2 & \text{ for } m=l\\
24(-1)^{m-l+1}z_0^6/(z_m-z_l)^4 & \text{ otherwise}
\end{array} \right.
\label{eq:Matrix_zz} 
\end{eqnarray}
These identities have been derived in \cite{Gor67,Good00}. The sign of the results for $m\ne l$ depends on the sign convention for the normalization constants in $\psi_m(z)$ and $\psi_l(z)$. Throughout this paper, we take the normalization constant to be real and positive for each state.

The numerical simulations in Figs. \ref{fig:StateEvolution} and \ref{fig:TransitionBothSpins} are obtained by solving these equations. Note that in both figures, we compare the numerical simulation for the population of the ground state to $H_0$, $\KlB{a_{1\uparrow}}^2+\KlB{a_{1\downarrow}}^2$, with the model prediction for the ground state of $H_{0,\pm}$, $\KlB{a_1^+}^2+\KlB{a_1^-}^2$. Due to the smallness of $\braket{1\uparrow|m\pm}$, $\braket{1\downarrow|m\pm}$ for $m > 1$, the different meaning of these populations is not visible.



\begin{thebibliography}{99}

\def\Lit#1#2#3#4{#1 {\bf #2}, #3 (#4)}
\def\LSJETP#1#2#3{\Lit{Soviet Physics JETP}{#1}{#2}{#3}}
\def\LPR#1#2#3{\Lit{Phys. Rev.}{#1}{#2}{#3}}
\def\ArXiv#1#2{arXiv:#1}
\def\LPRA#1#2#3{\Lit{Phys. Rev. A}{#1}{#2}{#3}}
\def\LPRB#1#2#3{\Lit{Phys. Rev. B}{#1}{#2}{#3}}
\def\LPRC#1#2#3{\Lit{Phys. Rev. C}{#1}{#2}{#3}}
\def\LPRD#1#2#3{\Lit{Phys. Rev. D}{#1}{#2}{#3}}
\def\LPRL#1#2#3{\Lit{Phys. Rev. Lett.}{#1}{#2}{#3}}
\def\LPhysToday#1#2#3{\Lit{Physics Today}{#1}{#2}{#3}}
\def\LJETPL#1#2#3{\Lit{JETP Lett.}{#1}{#2}{#3}}
\def\LPL#1#2#3{\Lit{Phys. Lett.}{#1}{#2}{#3}}
\def\LAPB#1#2#3{\Lit{Appl. Phys. B}{#1}{#2}{#3}}
\def\LAPParis#1#2#3{\Lit{Ann. Phys. (Paris)}{#1}{#2}{#3}}
\def\LJPG#1#2#3{\Lit{Journ. Phys. G}{#1}{#2}{#3}}
\def\LEPJC#1#2#3{\Lit{Europ. Phys. Journ. C}{#1}{#2}{#3}}
\def\LNat#1#2#3{\Lit{Nature}{#1}{#2}{#3}}
\def\LNC#1#2#3{\Lit{Il Nuovo Cimento}{#1}{#2}{#3}}
\def\LNatP#1#2#3{\Lit{Nat. Phys.}{#1}{#2}{#3}}
\def\LNIMA#1#2#3{\Lit{Nucl. Inst. Meth A}{#1}{#2}{#3}}
\def\LPRep#1#2#3{\Lit{Phys. Rep.}{#1}{#2}{#3}}
\def\LPLB#1#2#3{\Lit{Phys. Lett. B}{#1}{#2}{#3}}
\def\LCQG#1#2#3{\Lit{Class. Quant. Grav.}{#1}{#2}{#3}}
\def\LPPNP#1#2#3{\Lit{Prog. Part. Nucl. Phys.}{#1}{#2}{#3}}
\def\LARNPS#1#2#3{\Lit{Annu. Rev. Nucl. Part. Sci.}{#1}{#2}{#3}}
\def\LCRP#1#2#3{\Lit{Compt. Rend. Phys.}{#1}{#2}{#3}}
\def\LPU#1#2#3{\Lit{Phys.-Uspekhi}{#1}{#2}{#3}}
\def\LJCP#1#2#3{\Lit{J. Chem. Phys.}{#1}{#2}{#3}}
\def\LAJP#1#2#3{\Lit{Am. J. Phys.}{#1}{#2}{#3}}
\def\LJAP#1#2#3{\Lit{Journ. of Appl. Phys.}{#1}{#2}{#3}}
\def\LRSI#1#2#3{\Lit{Rev. Sci. Instr.}{#1}{#2}{#3}}
\def\LAHEP#1#2#3{\Lit{Adv. High En. Phys.}{#1}{#2}{#3}}

\bibitem{NesNat02} V.V. Nesvizhevsky, H.G. B\"orner, A.K. Petukhov, H. Abele, S. Bae\ss{}ler, F.J. Rue\ss{}, T. St\"oferle, A. Westphal, A.M. Gagarski, G.A. Petrov, A.V. Strelkov, \LNat{415}{297}{2002}.

\bibitem{NesPRD03} V.V. Nesvizhevsky, H.G. B\"orner, A.M. Gagarski, A.K. Petukhov, G.A. Petrov, H. Abele, S. Bae\ss{}ler, G. Divcovic, F.J. Rue\ss{}, T. St\"oferle, A. Westphal, A.V. Strelkov, K.V. Protasov, A.Yu. Voronin, \LPRD{67}{102002}{2003}. 

\bibitem{NesEPJC05} V.V. Nesvizhevsky, A.K. Petukhov, H.G. B\"orner, T.A. Baranova, A.M. Gagarski, G.A. Petrov, K.V. Protasov, A.Yu. Voronin, S. Bae\ss{}ler, H. Abele, A. Westphal, L. Lucovac, \LEPJC{40}{479}{2005}.

\bibitem{BaeJPG09} S. Bae\ss{}ler, \LJPG{36}{10}{2009}.

\bibitem{NesUsp10} V.V. Nesvizhevsky, \LPU{53}{645}{2010}. 

\bibitem{NesBook05} V. Nesvizhevsky, K. Protasov, in: D. Moore (Ed.), Trends in Quantum Gravity Research, Nova Science Publishers, New York, 2005, pp. 65-107.

\bibitem{JenNat11} T. Jenke, P. Geltenbort, H. Lemmel, H. Abele, \LNatP{7}{468}{2011}.

\bibitem{JenPRL14} T. Jenke, G. Cronenberg, J. Burgdörfer, L.A. Chizhova, P. Geltenbort, A.N. Ivanov, T. Lauer, T. Lins, S. Rotter, H. Saul, U. Schmidt, H. Abele, \LPRL{112}{151105}{2014}.

\bibitem{Kre09} M. Kreuz, V.V. Nesvizhevsky, P. Schmidt-Wellenburg, T. Soldner, M. Thomas, H.G. B\"orner, F. Naraghi, G. Pignol, K.V. Protasov, D. Rebreyend, F. Vezzu, D. Forest, P. Ganau, J.M. Mackowski, C. Michel, J.L. Montorio, N. Morgado, L. Pinard, A. Remillieux, S. Bae\ss{}ler, A.M. Gagarski, L.A. Grigorieva, A.M. Kuzmina, A.E. Meyerovich, L.P. Mezhov-Deglin, G.A. Petrov, A.V. Strelkov, A.Yu. Voronin, \LNIMA{611}{326}{2009}.

\bibitem{Pig09} G. Pignol, Thesis, Université Joseph Fourier, Grenoble (2009).

\bibitem{Bae11a} S. Bae\ss{}ler, M. Beau, M. Kreuz, V. N. Kurlov, V.V. Nesvizhevsky, G. Pignol, K.V. Protasov, F. Vezzu, A.Yu. Voronin, \LCRP{12}{707}{2011}.

\bibitem{Nes07} V.V. Nesvizhevsky, A.K. Petukhov, H.G. B\"orner, T. Soldner, P. Schmidt-Wellenburg, M. Kreuz, G. Pignol, K.V. Protasov, D. Rebreyend, F. Vezzu, D. Forest, P. Ganau, J.M. Mackowski, C. Michel, J.L. Montorio, N. Morgado, L. Pinard, A. Remillieux, A.M. Gagarski, G.A. Petrov, A.M. Kusmina, A.V. Strelkov, H. Abele, S. Bae\ss{}ler, A.Yu. Voronin, 
Proceedings of "15th International Seminar on Interaction of Neutrons with Nuclei", Dubna, Russia (2007) (\ArXiv{0708.2541}{2007}).

\bibitem{Ant11} I. Antoniadis, S. Baeßler, M. Büchner, V.V. Fedorov, S. Hoedl, A. Lambrecht, V.V. Nesvizhevsky, G. Pignol, K.V. Protasov, S. Reynaud, Yu. Sobolev, \LCRP{12}{755}{2011}.

\bibitem{Bra11} P. Brax, G. Pignol, \LPRL{107}{2011}{111301}.

\bibitem{Pet99} A. Peters, K. Chung, S. Chu, \LNat{400}{849}{1999}.

\bibitem{Adel09} E.G. Adelberger, J.H. Gundlach, B.R. Heckel, S. Hoedl, S. Schlamminger, \LPPNP{62}{102}{2009}.

\bibitem{Nes00} V.V. Nesvizhevsky, H. B\"orner, A.M. Gagarski, G.A. Petrov, A.K. Petukhov, H. Abele, S. Bae\ss{}ler, T. St\"oferle, S.M. Soloviev, \LNIMA{440}{754}{2000}.

\bibitem{Pign14} G. Pignol, S. Baeßler, V.V. Nesvizhevsky, K. Protasov, D. Rebreyend, A. Voronin, \LAHEP{2014}{628125}{2014}.

\bibitem{Abe10} H. Abele, T. Jenke, H. Leeb, J. Schmiedmayer, \LPRD{81}{065019}{2010}.









\bibitem{Shir63} J.H. Shirley, \LJAP{34}{783}{1963}.

\bibitem{Blo40} F. Bloch, A. Siegert, \LPR{57}{522}{1940}.




\bibitem{Mey02} A. E. Meyerovich, I. V. Ponomarev, \LPRB{65}{155413}{2002}.

\bibitem{Vor06} A.Yu. Voronin, H. Abele, S. Bae\ss{}ler, V.V. Nesvizhevsky, A. K. Petukhov, K.V. Protasov, A. Westphal, \LPRD{73}{044029}{2006}.

\bibitem{Shir65} J.H. Shirley, \LPR{138}{979}{1965}.

\bibitem{Good00} D.M. Goodmanson, \LAJP{68}{866}{2000}.

\bibitem{Gor67} R.J. Gordon, \LJCP{51}{14}{1969}.

\end{thebibliography}
\end{document}